\documentclass[twoside]{article}

\usepackage[a4paper]{geometry}
\usepackage[latin1]{inputenc} 
\usepackage[T1]{fontenc} 
\usepackage{RR}
\usepackage{hyperref}
\usepackage{graphicx}
\usepackage{epsfig}
\usepackage{subfigure}
\usepackage{amsmath,amsgen,amstext,amssymb}
\usepackage{amsfonts}
\usepackage{latexsym}
\usepackage{bbm}

\newtheorem{lemma}{Lemma}[section]
\newtheorem{proposition}{Proposition}[section]

\newtheorem{remark}{Remark}[section]

\def\done{\hspace*{\fill} \rule{1.8mm}{2.5mm}}

\RRdate{June 2012}

\RRauthor{
Konstantin Avrachenkov\thanks{Inria Sophia Antipolis, France, K.Avrachenkov@sophia.inria.fr}%
\and
Philippe Nain\thanks{Inria Sophia Antipolis, France, Philippe.Nain@inria.fr}%
\and
Uri Yechiali\thanks{Tel Aviv University, Israel, uriy@post.tau.ac.il}
}

\authorhead{K. Avrachenkov \& P. Nain \& U. Yechiali}

\RRtitle{Files d'attente avec r\'e\'emissions des clients et un serveur unique}

\RRetitle{A retrial system with two input streams\\ and two orbit queues}

\titlehead{Files d'attente avec r\'e\'emissions des clients}

\RRresume{Un serveur unique 'exponentiel', sans file d'attente, est aliment\'e par deux flux
poissonniens et  ind\'ependants de clients.  Un client du flux $i$
($i=1,2$) trouvant le serveur unique inoccup\'e est servi et quitte le syst\`eme une fois servi;
s'il trouve le serveur unique occup\'e alors il rejoint  une file d'attente,
appel\'ee orbite $i$,  de capacit\'e illimit\'ee et  \'equip\'ee d'un serveur 'exponentiel'.
Un client quittant l'orbite $i$  est rout\'e vers le serveur unique; si celui-ci
est inoccup\'e alors il quitte le syst\`eme une fois servi; s'il est occup\'e alors il est de nouveau
rout\'e vers l'orbite $i$ et le processus d\'ecrit ci-dessus se r\'ep\`ete.
Le tout cr\'e\'e ainsi un syst\`eme compos\'e de trois file d'attente fortement corr\'el\'ees.  Ce syst\`eme
de files d'attente sert \`a modeliser des protocoles de communication \`a contention.
Nous \'etablissons une \'equation satisfaite par la fonction g\'en\'eratrice jointe du nombre
de clients dans les trois files d'attente, que nous r\'esolvons par r\'eduction \`a un probl\`eme
aux limites de Riemann-Hilbert. Nous d\'eterminons la condition de stabilit\'e du syst\`eme
et concluons le rapport en pr\'esentant des r\'esultats num\'eriques pour les principales mesures
de performance.}

\RRabstract{Two independent Poisson streams of jobs flow into a single-server service system having a limited common buffer
that can hold at most one job. If a type-$i$ job ($i=1,2$) finds the server busy, it is blocked and routed to a separate
type-$i$ retrial (orbit) queue that attempts to re-dispatch its jobs at its specific Poisson rate. This creates a system
with three dependent queues. Such a queueing system serves as a model for two competing job streams in a carrier sensing
multiple access system. We study the queueing system using multi-dimensional probability generating functions, and derive
its necessary and sufficient stability conditions while solving a boundary value problem. Various performance measures
are calculated and numerical results are presented.}

\RRmotcle{Files d'attente r\'e\'emissions des clients, Probl\`eme aux limites de Riemann-Hilbert,
Protocoles de communication \`a contention}

\RRkeyword{Retrial queues, Riemann-Hilbert boundary value problem, Carrier sensing
multiple access system}

\RRprojet{Maestro}

\RCSophia

\begin{document}

\RRNo{7999}
\makeRR

\section{Introduction}

Queues with blocking and with retrials have been studied extensively in the literature
(see e.g. \cite{A99}-\cite{AMNS12}, \cite{BY91}-\cite{CRP93}, \cite{FT97}, \cite{F86}, \cite{Y88} and references therein).
In this paper we investigate a single-server system with two independent exogenous Poisson streams flowing into a common buffer
that can hold at most one job. If a type-$i$ job finds the server busy, it is routed to a separate retrial (orbit) queue from which jobs are re-transmitted at a Poisson rate. Such a queueing system serves as a model for two competing job streams in a carrier sensing
multiple access system, where the jobs -- after a failed attempt to network access -- wait in an orbit queue \cite{Nain85, S94}.
The two types of customers can be interpreted as customers with different priority requirements. An important feature of the
retrial system under consideration is a constant retrial rate. The constant retrial rate helps to stabilize the multiple
access system \cite{BG92}. The retrial queueing systems with a constant retrial rate and
a single type of jobs has been considered in \cite{AGN01}-\cite{AMNS12}, \cite{CPP93}-\cite{CRP93}, \cite{F86}.
We formulate this system as a three-dimensional Markovian queueing network,
and derive its necessary and sufficient stability conditions. Recently these stability conditions have been shown
by simulations to hold even in a more general system with generally distributed service times \cite{AMNS12}.

The structure of the paper is as follows:
After the Introduction we present the model in Section~2. Balance equations and generating functions are derived in Section~3, while necessary
stability conditions are obtained in Section~4. Using the technique developed by Fayolle and Iasnogoroski \cite{FI79}, in Section~5
we show that these generating functions
are obtained, in closed-form, via the solution of a Riemann-Hilbert boundary value problem. This approach allows us to show that the necessary stability conditions found in Section~4 are also sufficient.
Performance measures are calculated in Section~6, and numerical results are presented in Section~7.
In particular, our numerical results demonstrate that the proposed multiple access system with two types of jobs and constant
retrial rates provides incentives for the users to respect the contracts.

\section{Model}

\begin{figure}[t]
\centering
\includegraphics[width=8cm]{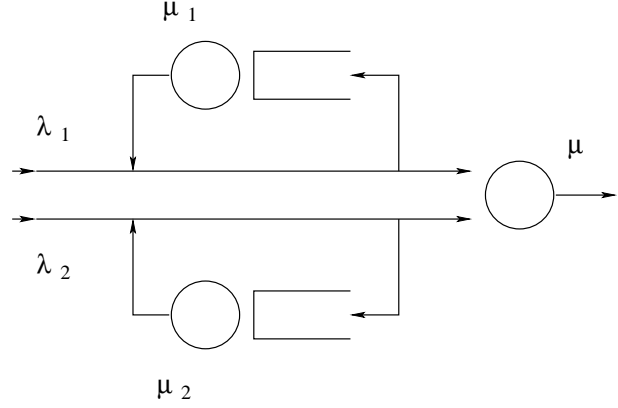}
\caption{Retrial system with two orbit queues.}
\label{fig:model}
\end{figure}

Two independent Poisson streams of jobs, $S_1$ and $S_2$, flow into a single-server
service system. The service system can hold {\em at most} one job. The arrival rate of stream
$S_i$ is $\lambda_i$, $i=1,2$, with $\lambda:=\lambda_1+\lambda_2$. The required service time
of each job is independent of its type and is exponentially distributed with mean $1/\mu$.
If an arriving type-$i$ job finds the (main) server busy, it is routed to a dedicated retrial
(orbit) queue that operates as an $\cdot/M/1/\infty$ queue. That is, blocked jobs of type $i$
form a type-$i$ single-server orbit queue that attempts to retransmit jobs (if any) to the
main service system at a Poisson rate of $\mu_i$, $i=1,2$.
Thus, the overall system is comprised of three queues as depicted in Figure~\ref{fig:model}.

\section{Balance equations and generating functions}

\begin{figure}[ht]
\centering
\includegraphics[width=12cm]{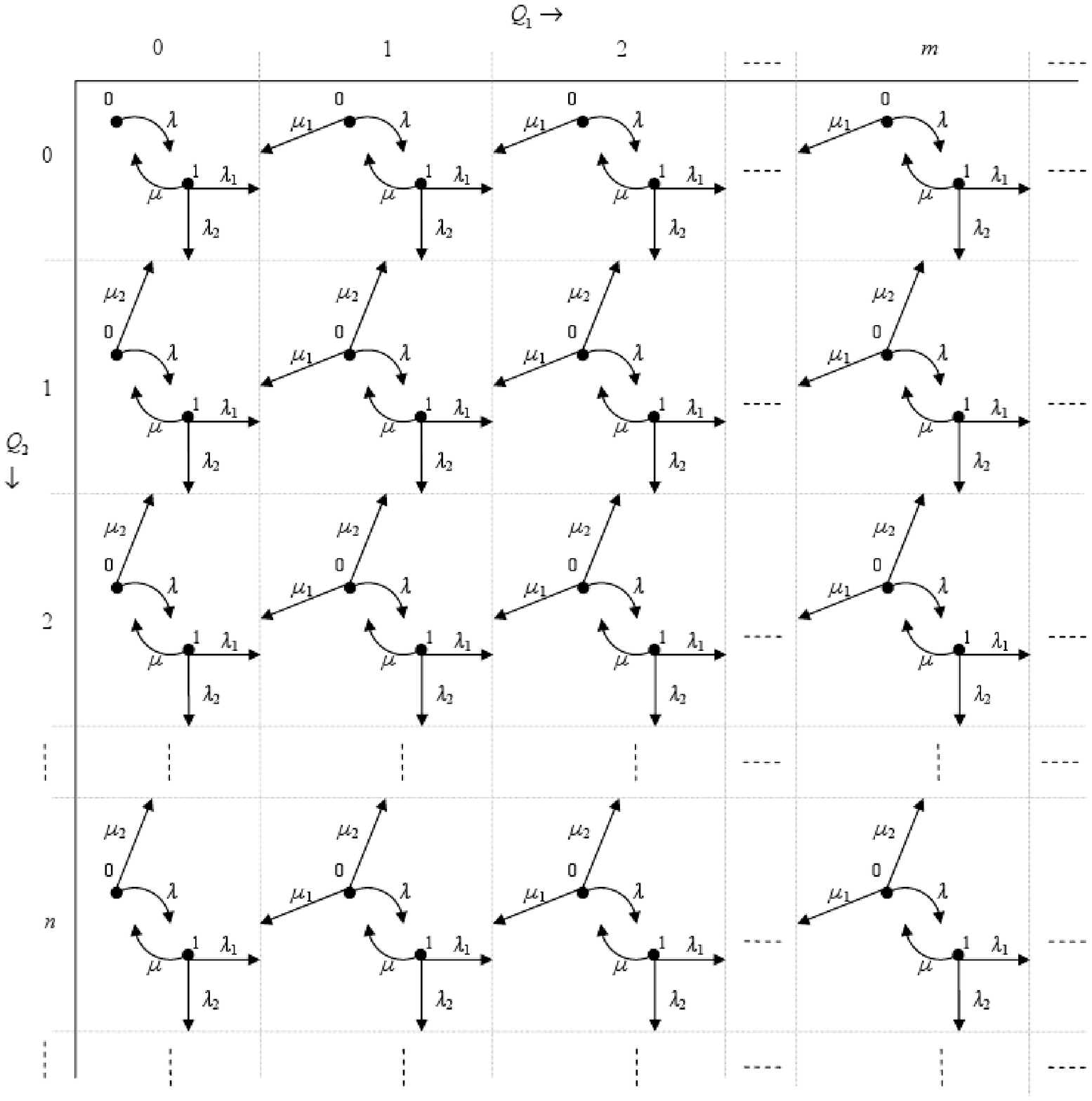}
\caption{Transition-rate diagram.}
\label{fig:RateDiagram}
\end{figure}

Consider the system in steady state.
Let $L$ denote the number of jobs in the main queue. $L$ assumes the values of 0 or 1.
Let $Q_i$ be the number of jobs in orbit-queue $i$, $i=1,2$. The transition-rate
diagram of the system is depicted in Figure~\ref{fig:RateDiagram}. The numbers 0 or 1
appearing next to each node indicate whether $L=0$ or $L=1$, respectively.

Define the set of stationary probabilities $\{P_{mn}(k)\}$ as follows:
$$
P_{mn}(k) = P(Q_1=m,Q_2=n,L=k),
\quad m,n=0,1,2,... \quad k=0,1.
$$
Define the marginal probabilities
$$
P_{m\bullet}(k)=\sum_{n=0}^\infty P_{mn}(k) = P(Q_1=m,L=k),
\quad m=0,1,2,... \quad k=0,1,
$$
and
$$
P_{\bullet n}(k)=\sum_{m=0}^\infty P_{mn}(k) = P(Q_2=n,L=k),
\quad n=0,1,2,... \quad k=0,1.
$$

Let us write the balance equations.
If $Q_2=0$, we have
\begin{itemize}
\item[(a)] for $Q_1=0$ and $k=0$,
\begin{equation}\label{eq:Q20Q10k0}
\lambda P_{00}(0) = \mu P_{00}(1),
\end{equation}
\item[(b)] for $Q_1=m \ge 1$ and $k=0$,
\begin{equation}\label{eq:Q20Q1mk0}
(\lambda+\mu_1) P_{m0}(0) = \mu P_{m0}(1),
\end{equation}
\item[(c)] for $Q_1=0$ and $k=1$,
\begin{equation}\label{eq:Q20Q10k1}
(\lambda+\mu) P_{00}(1) = \lambda P_{00}(0) + \mu_1 P_{10}(0) + \mu_2 P_{01}(0),
\end{equation}
\item[(d)] for $Q_1=m \ge 1$ and $k=1$,
\begin{equation}\label{eq:Q20Q1mk1}
(\lambda+\mu) P_{m0}(1) = \lambda P_{m0}(0) + \mu_1 P_{m+1,0}(0)
+ \mu_2 P_{m1}(0) + \lambda_1 P_{m-1,0}(1).
\end{equation}
\end{itemize}

\medskip

\noindent
If $Q_2=n$, $n \ge 1$, we have
\begin{itemize}
\item[(e)] for $Q_1=0$ and $k=0$,
\begin{equation}\label{eq:Q2nQ10k0}
(\lambda + \mu_2) P_{0n}(0) = \mu P_{0n}(1),
\end{equation}
\item[(f)] for $Q_1=m \ge 1$ and $k=0$,
\begin{equation}\label{eq:Q2nQ1mk0}
(\lambda + \mu_1 + \mu_2) P_{mn}(0) = \mu P_{mn}(1),
\end{equation}
\item[(g)] for $Q_1=0$ and $k=1$,
\begin{equation}\label{eq:Q2nQ10k1}
(\lambda + \mu) P_{0n}(1) = \lambda P_{0n}(0) + \mu_1 P_{1n}(0) +\mu_2 P_{o,n+1}(0)+\lambda_2 P_{0,n-1}(1),
\end{equation}
\item[(h)] for $Q_1=m \ge 1$ and $k=1$,
$$
(\lambda + \mu) P_{mn}(1) = \lambda P_{mn}(0) + \mu_1 P_{m+1,n}(0) + \mu_2 P_{m,n+1}(0)
$$
\begin{equation}\label{eq:Q2nQ1mk1}
+\lambda_1 P_{m-1,n}(1) +\lambda_2 P_{m,n-1}(1).
\end{equation}
\end{itemize}

Let us define the following Probability Generating Functions (PGFs):
$$
G_n^{(k)}(x) = \sum_{m=0}^\infty P_{mn}(k) x^m,
\quad k=0,1, \quad n \ge 0.
$$
Then, for $n=0$ and $k=0$, multiplying each equation from (\ref{eq:Q20Q10k0})
and (\ref{eq:Q20Q1mk0}) by $x^m$, respectively, and summing over $m$ results in
$$
\lambda \sum_{m=0}^\infty P_{m0}(0) x^m + \mu_1 \sum_{m=1}^\infty P_{m0}(0) x^m
= \mu \sum_{m=0}^\infty P_{m0}(1) x^m,
$$
or
\begin{equation}\label{eq:G1}
(\lambda+\mu_1) G_0^{(0)}(x) - \mu_1 P_{00}(0) = \mu G_0^{(1)}(x).
\end{equation}
Similarly, for $n=0$ and $k=1$, using equations (\ref{eq:Q20Q10k1}) and (\ref{eq:Q20Q1mk1})
leads to
$$
(\lambda+\mu) G_0^{(1)} = \lambda G_0^{(0)} + \mu_1 \sum_{m=0}^\infty P_{m+1,0}(0) x^m
+\mu_2G_1^{(0)}(x)+\lambda_1 \sum_{m=1}^\infty P_{m-1,0}(1) x^m.
$$
That is,
$$
(\lambda+\mu) G_0^{(1)}(x) = \lambda G_0^{(0)}(x) + \frac{\mu_1}{x}(G_0^{(0)}(x)-P_{00}(0))
+\mu_2 G_1^{(0)}(x) + \lambda_1 x G_0^{(1)}(x).
$$
Multiplying by $x$ and arranging terms, we obtain
\begin{equation}\label{eq:G2}
-(\lambda x+\mu_1) G_0^{(0)}(x) + (\lambda_1 (1-x)+\lambda_2+\mu)xG_0^{(1)}(x)
-\mu_2 x G_1^{(0)}(x) = -\mu_1 P_{00}(0).
\end{equation}
Using equations (\ref{eq:Q2nQ10k0}) and (\ref{eq:Q2nQ1mk0}) for $n \ge 1$ and $k=0$
results in
$$
(\lambda+\mu_2) G_n^{(0)}(x) + \mu_1 (G_n^{(0)}(x)-P_{0n}(0)) = \mu G_n^{(1)}(x),
$$
or
\begin{equation}\label{eq:G3}
(\lambda+\mu_1+\mu_2)G_n^{(0)}(x) - \mu G_n^{(1)}(x) = \mu_1 P_{0n}(0).
\end{equation}
Similarly, for $n \ge 1$ and $k=1$, equations (\ref{eq:Q2nQ10k1}) and (\ref{eq:Q2nQ1mk1})
lead to
$$
(\lambda+\mu)G_n^{(1)}(x) = \lambda G_n^{(0)}(x) + \frac{\mu_1}{x}(G_n^{(0)}(x)-P_{0n}(0))
+\mu_2 G_{n+1}^{(0)}(x)
$$
$$
+ \lambda_1 x G_n^{(1)}(x) + \lambda_2 G_{n-1}^{(1)}(x),
$$
or
$$
-(\lambda x+\mu_1)G_n^{(0)}(x)+(\lambda_1(1-x)+\lambda_2+\mu)xG_n^{(1)}(x)-\mu_2 x G_{n+1}^{(0)}(x)
$$
\begin{equation}\label{eq:G4}
-\lambda_2 x G_{n-1}^{(1)}(x) = - \mu_1 P_{0n}(0).
\end{equation}
Define now the two-dimensional PGFs
\begin{equation}
\label{def:Hxy}
H^{(k)}(x,y) = \sum_{n=0}^\infty \sum_{m=0}^\infty P_{mn}(k) x^m y^n
= \sum_{n=0}^\infty G_n^{(k)}(x) y^n, \quad k=0,1.
\end{equation}
Using equations (\ref{eq:G1}) and (\ref{eq:G3}), multiplying respectively by $y^n$ and summing
over $n$, we obtain
\begin{equation}\label{eq:H0}
(\lambda+\mu_1)H^{(0)}(x,y)+\mu_2(H^{(0)}(x,y)-G_0^{(0)}(x))-\mu H^{(1)}(x,y) = \mu_1 H^{(0)}(0,y).
\end{equation}
Similarly, using equations (\ref{eq:G2}) and (\ref{eq:G4}), we obtain
$$
-(\lambda x + \mu_1)H^{(0)}(x,y) + (\lambda_1(1-x)+\lambda_2+\mu)x H^{(1)}(x,y)
$$
\begin{equation}\label{eq:H1}
-\frac{\mu_2 x}{y}(H^{(0)}(x,y)-G_0^{(0)}(x))-\lambda_2 xy H^{(1)}(x,y) = - \mu_1 H^{(0)}(0,y).
\end{equation}

Noting that $G_0^{(0)}(x)=H^{(0)}(x,0)$ and denoting $\alpha:=\lambda+\mu_1+\mu_2$,
we can rewrite equations (\ref{eq:H0}) and (\ref{eq:H1}) as
\begin{eqnarray}
\alpha H^{(0)}(x,y)-\mu H^{(1)}(x,y)&=&\mu_2 H^{(0)}(x,0)+\mu_1 H^{(0)}(0,y),
\label{eq1}\\
(\lambda x y +\mu_1y+\mu_2 x) H^{(0)}(x,y) &-& (\lambda_1 (1-x) +\lambda_2 (1-y)+\mu) x y H^{(1)}(x,y)\nonumber\\
&=&\mu_2 x H^{(0)}(x,0)+\mu_1 y H^{(0)}(0,y), \label{eq2}
\end{eqnarray}
or, equivalently, in a matrix form
\begin{equation}\label{eq:Mform}
{\bf C}(x,y) {\bf H}(x,y) = {\bf g}(x,y),
\end{equation}
where
$$
{\bf C}(x,y)=
\left[\begin{array}{cc}
\alpha & -\mu \\
\lambda x y + \mu_1 y + \mu_2 x & -(\lambda_1(1-x)+\lambda_2(1-y)+\mu)xy
\end{array}\right],
$$
$$
{\bf H}(x,y)=
\left[\begin{array}{c}
H^{(0)}(x,y) \\
H^{(1)}(x,y)
\end{array}\right],
$$
$$
{\bf g}(x,y)=
\left[\begin{array}{c}
\mu_2 H^{(0)}(x,0) + \mu_1 H^{(0)}(0,y) \\
\mu_2 x H^{(0)}(x,0) + \mu_1 y H^{(0)}(0,y)
\end{array}\right].
$$
Now, if we calculate $H^{(0)}(x,0)$ and $H^{(0)}(0,y)$, the two-dimensional PGF
${\bf H}(x,y)$ is immediately obtained from equation (\ref{eq:Mform}).

\section{Necessary stability conditions}
\label{sec:necessary}

\begin{proposition}
\label{prop:necessary}
\begin{equation}
\label{eq:PL1}
H^{(1)}(1,1)=P(L=1)=\frac{\lambda}{\mu}
\end{equation}
and
\begin{eqnarray}
\label{eq:109-1}
H^{(0)}(0,1)&=&P(Q_1=0,L=0)= 1 - \frac{\lambda}{\mu} \left(1+\frac{\lambda_1}{\mu_1}\right)\\
H^{(0)}(1,0)&=&P(Q_2=0,L=0)= 1 - \frac{\lambda}{\mu} \left(1+\frac{\lambda_2}{\mu_2}\right).
\label{eq:109-2}
\end{eqnarray}
\end{proposition}

The identities (\ref{eq:PL1})-(\ref{eq:109-2}) show that conditions
(i) $\lambda/\mu\leq 1$ and (ii) $(\lambda/\mu)(1+\lambda_i/\mu_i)\leq 1$ for $i=1,2$,
are necessary for the existence of a steady-state. Note that (i) is a consequence of (ii) so that in the following
we will not consider condition (i) but only conditions (ii).

{\bf Proof of Proposition \ref{prop:necessary}:}
For each $m=0,1,2,...$ we consider a vertical ``cut'' (see Figure~\ref{fig:RateDiagram}) between the column
representing the states $\{Q_1=m,L=1\}$ and the column representing the states $\{Q_1=m+1,L=0\}$.
According to the local balance equation approach \cite{Bocharov04}, we can write the balance of rates
between the states from the left of the cut and the states from the right of the cut. Namely, we have
\begin{equation}\label{eq:102}
\lambda_1 P_{m\bullet}(1) = \mu_1 P_{m+1\bullet}(0),
\quad m=0,1,2,... .
\end{equation}
Summing (\ref{eq:102}) over all $m$ results in
\begin{equation}\label{eq:103}
\lambda_1 H^{(1)}(1,1) = \mu_1 (1-H^{(1)}(1,1)-P_{0\bullet}(0)).
\end{equation}
Clearly, $P(L=k)=\sum_{m=0}^\infty P_{m \bullet}(k)=H^{(k)}(1,1)$, $k=0,1$.

\medskip

From (\ref{eq:103}) we readily get
\begin{equation}\label{eq:104}
1-P_{0\bullet}(0) = \frac{\lambda_1+\mu_1}{\mu_1} H^{(1)}(1,1).
\end{equation}
Since $P_{0\bullet}(0)=H^{(0)}(0,1)$, we can write (\ref{eq:104}) as
\begin{equation}\label{eq:105}
1-H^{(0)}(0,1) = \frac{\lambda_1+\mu_1}{\mu_1} H^{(1)}(1,1),
\end{equation}
and, by symmetry,
\begin{equation}\label{eq:106}
1-H^{(0)}(1,0)= \frac{\lambda_2+\mu_2}{\mu_2} H^{(1)}(1,1).
\end{equation}
Substituting (\ref{eq:105}) and (\ref{eq:106}) in equation (\ref{eq1}), with $x=y=1$,
yields
$$
H^{(1)}(1,1)=P(L=1)=\frac{\lambda}{\mu}.
$$
Now, from (\ref{eq:105}) and (\ref{eq:106}), respectively, we obtain
\begin{equation}\label{eq:108}
H^{(0)}(0,1) = P(Q_1=0,L=0) = 1 - \frac{\lambda}{\mu}\left(\frac{\lambda_1+\mu_1}{\mu_1}\right)
\end{equation}
and
\begin{equation}\label{eq:109}
H^{(0)}(1,0) = P(Q_2=0,L=0) = 1 - \frac{\lambda}{\mu}\left(\frac{\lambda_2+\mu_2}{\mu_2}\right),
\end{equation}
which completes the proof.
\hfill\done

The next result shows that the system cannot be stable if either  $(\lambda/\mu)(1+\lambda_1/\mu_1)= 1$
or $(\lambda/\mu)(1+\lambda_2/\mu_2)= 1$.

\begin{proposition}
\label{prop:necessary-2}
If either  $(\lambda/\mu)(1+\lambda_1/\mu_1)= 1$ or  $(\lambda/\mu)(1+\lambda_2/\mu_2)= 1$ then
$P_{m,n}(0)=P_{m,n}(1)=0$ for all $m,n=0,1,\ldots $ or, equivalently, both queues $Q_1$ and $Q_2$ are unbounded
with probability one.
\end{proposition}

{\bf Proof.}
Assume, for instance, that  $(\lambda/\mu)(1+\lambda_2/\mu_2)= 1$ so that $H^{(0)}(1,0)=0$ from
(\ref{eq:109-2}). Since   $H^{(0)}(1,0)=\sum_{m\geq 0} P_{m, 0}(0)$ (see (\ref{def:Hxy})), the condition $H^{(0)}(1,0)=0$ implies
that
\begin{equation}
 P_{m, 0}(0)=0 \quad \hbox{for } m=0,1,\ldots,
\label{eq:1000}
\end{equation}
so that from (\ref{eq:Q20Q10k0})-(\ref{eq:Q20Q1mk0})
\begin{equation}
 P_{m, 0}(1)=0 \quad \hbox{for } m=0,1,\ldots \ .
\label{eq:1001}
\end{equation}

We now use an induction argument to prove that
\begin{equation}
\label{induction}
P_{m, n}(0)=0 \quad \hbox{for } m,n=0,1,\ldots.
\end{equation}
We have already shown in (\ref{eq:1000}) that (\ref{induction})
is true for $n=0$. Assume that (\ref{induction}) is true for $n=0,1,\ldots,k$ and let us show that it is still
true for $n=k+1$.


From (\ref{eq:Q2nQ1mk0}) and the induction
hypothesis we get that $P_{m,k}(0)= P_{m,k}(1) = 0$ for $m= 1,2,\ldots$.
The latter equality implies, using (\ref{eq:Q2nQ1mk1}), that  $P_{m,k+1}(0)= 0$. This shows
that (\ref{induction}) holds for $m=0,1,...$ and $n=k+1$, and completes the induction argument, proving
that (\ref{induction}) is true.

We have therefore proved that $P_{m, n}(0)=0$ for all $m,n=0,1,\ldots$. Let us prove that
$P_{m ,n}(1)=0$ for all $m,n=0,1,\ldots$. The latter is true for $m,n=1,2,\ldots$
thanks to (\ref{eq:Q2nQ1mk0}). It is also true for $n=0$, $m=0,1,\ldots$ from (\ref{eq:1001}).
It remains to investigate the case where $m=0$ and $n=0,1,\ldots$. By (\ref{eq:Q2nQ10k0})
and (\ref{induction}) we get that $P_{0,n}(1)=0$ for $n=1,2,\ldots$, whereas we have already noticed that
$P_{0,0}(1)=0$.

In summary,  $P_{m, n}(0)=P_{m, n}(1)=0$ for all $m,n=0,1,\ldots$, so that
\[
P(Q_1=m,Q_2=n)=P_{m, n}(0)+P_{m, n}(1)=0
\]
for all $m,n=0,1,\ldots$, which completes the proof.
\hfill\done

We conclude from Propositions (\ref{prop:necessary}) and (\ref{prop:necessary-2}) that conditions
\begin{equation}
\label{stability}
\left(\frac{\lambda}{\mu}\right)\left(1+\frac{\lambda_1}{\mu_1}\right)<1
\quad \hbox{and} \quad
\left(\frac{\lambda}{\mu}\right)\left(1+\frac{\lambda_2}{\mu_2}\right)<1
\end{equation}
are necessary for the system to be stable. We will show in Section \ref{sec:BVP} that these conditions
are also sufficient, thereby implying that they are the stability conditions of the system.

\section{Derivation of $H^{(0)}(x,0)$ and $H^{(1)}(0,y)$}
\label{sec:BVP}

Throughout we assume that the necessary stability conditions found in (\ref{stability}) hold.
Our analysis below will formally show that these conditions are also sufficient
for the stability of the system. Let us give an intuitive motivation for this result.
In a stable system,
$\lambda/\mu$ is the fraction of time the server in the main queue is busy. Thus, this is also
the proportion of jobs sent to the orbit queues. Therefore, the maximal rates at which jobs flow into
orbit queue 1 and into orbit queue 2 are $(\lambda_1+\mu_1)\lambda/\mu$ and $(\lambda_2+\mu_2)\lambda/\mu$,
respectively. Each of these rates must be smaller than the corresponding maximal service rate, $\mu_1$ or $\mu_2$,
respectively.

\begin{lemma}
\label{lem:stab}
Conditions (\ref{stability}) imply that either $\alpha\lambda_1<\mu\mu_1$ or $\alpha\lambda_2<\mu\mu_2$.
\end{lemma}

{\bf Proof:}
Assume that  $\alpha\lambda_1\geq \mu\mu_1$ and  $\alpha\lambda_2 \geq \mu\mu_2$

Multiplying the first inequality in (\ref{stability}) by $\mu\mu_1$ and using the definition of
$\lambda$ and $\alpha$ gives
\[
(\lambda_1+\lambda_2)(\lambda_1+\mu_1)< \mu \mu_1\leq \alpha\lambda_1=(\lambda_1+\lambda_2+\mu_1+\mu_2)\lambda_1
\]
which is true if and only if (a) $\lambda_2\mu_1<\lambda_1 \mu_2$.

Multiplying now the second inequality in (\ref{stability}) by $\mu\mu_2$ gives
\[
(\lambda_1+\lambda_2)(\lambda_2+\mu_2)< \mu \mu_2\leq \alpha\lambda_2=(\lambda_1+\lambda_2+\mu_1+\mu_2)\lambda_2
\]
which is true if and only if (b) $\lambda_1\mu_2<\lambda_2 \mu_1$.

Since inequalities (a) and (b) cannot be true simultaneously we conclude that either $\alpha\lambda_1<\mu\mu_1$
or $\alpha\lambda_2<\mu\mu_2$, which concludes the proof.
\hfill\done

From equations (\ref{eq1})-(\ref{eq2}) we obtain the two-dimensional functional equation
\begin{equation}
R(x,y) H^{(0)}(x,y)=A(x,y)H^{(0)}(x,0)+B(x,y)H^{(0)}(0,y),\quad |x|\leq 1, |y|\leq 1,
\label{funct-eq}
\end{equation}
with
\begin{eqnarray}
R(x,y) &:=& \lambda_1 \alpha(1-x) x y +\lambda_2\alpha  (1-y) x y -\mu \mu_1 (1-x)y- \mu\mu_2 (1-y)x
\label{def-K}\\
A(x,y)&:=&((1-y)(\lambda_2 y-\mu)+\lambda_1(1-x)y) \mu_2 x 
\label{def-A}\\
B(x,y)&: = &((1-x)(\lambda_1 x-\mu)+\lambda_2(1-y)x)\mu_1 y.
\label{def-B}
\end{eqnarray}
For further use note that
\begin{eqnarray}
R(x,y)&=&\frac{\alpha}{\mu_2} A(x,y)+\lambda\mu(1-y)x +\mu\mu_1(x-y),
\label{R-A}\\
R(x,y)&=&\frac{\alpha}{\mu_1} B(x,y)+\lambda\mu(1-x)y +\mu\mu_2(y-x).
\label{R-B}
\end{eqnarray}
The kernel $R(x,y)$ of the functional equation (\ref{funct-eq}) is the same as the kernel in \cite[Eq. (1.3)]{FI79}
upon replacing $\lambda_i$ and $\mu_i$ in \cite{FI79} by $\lambda_i \alpha$ and $\mu_i\mu$, respectively, for $i=1,2$.

In the following we set $\hat\lambda_i=\alpha \lambda_i$ and $\hat \mu_i=\mu \mu_i$ for $i=1,2$. In this notation, the kernel
$R(x,y)$ is expressed as
\begin{equation}
\label{def-K2}
R(x,y)= \hat \lambda_1(1-x) x y +\hat \lambda_2  (1-y) x y - \hat \mu_1 (1-x)y- \hat \mu_2 (1-y)x.
\end{equation}
Also define $\hat \lambda=\hat \lambda_1 + \hat \lambda_2=\alpha \lambda$.

\vskip0.2true cm
{\bf Assumption A:} Without loss of generality thanks to Lemma \ref{lem:stab}, we will assume throughout that
 $\alpha\lambda_1<\mu\mu_1$ or, equivalently, that $\hat \lambda_1<\hat\mu_1$.
\vskip0.2true cm
Once $H^{(0)}(x,y)$ is known for all $|x|\leq 1$ and $|y|\leq 1$ then $H^{(1)}(x,y)$ can be found from (\ref{eq1}).
In the following we will therefore only focus on the calculation of  $H^{(0)}(x,y)$ or, equivalently from (\ref{funct-eq}),
on the calculation of $H^{(0)}(x,0)$ and $H^{(0)}(0,y)$ for all  $|x|\leq 1$ and $|y|\leq 1$.

We will show in Section \ref{ssec:BVP} that $H^{(0)}(x,0)$ is given by  the solution of a Riemann-Hilbert problem on the circle centered at $x=0$ and
with radius $\sqrt{\hat \mu_1/\hat\lambda_1}$ (see (\ref{sol-bvp})),
from which we will derive $H^{(0)}(0,y)$ for all $|y|\leq 1$ (see (\ref{eq:H00y})).

The technique of reducing the solution of certain two-dimensional functional equations  (equation (\ref{funct-eq})
in our case) to the solution of a boundary value problem (typically Rieman-Hilbert or Dirichlet problem) --
whose solution is known in closed-form -- is due to Fayolle and Iasnogorodski \cite{FI79}.
In \cite{FI79} (see also \cite{FKM82} that generalizes the work in \cite{FI79}) the unknown function is the generating function of a
two-dimensional stationary Markov chain describing the joint
queue-length in a two-queue system.
Cohen and Boxma \cite{CB83} extended the work in \cite{FI79,FKM82} to two-dimensional stationary Markov chains taking real values,
typically representing the joint waiting time or the joint unfinished work in a variety of two-queue systems. Other related papers include \cite{Blanc84, BIN88, FIM83, Nain85} (non-exhaustive list).

\subsection{Branching roots of $R(x,y)$}

For $y$ fixed, $R(x,y)$ vanishes at
\begin{equation}
\label{def-xy}
x(y)= \frac{-b(y)\pm \sqrt{c(y)}} {2 \hat\lambda_1 y}
\end{equation}
where
\begin{eqnarray}
b(y) &:=&\hat \lambda_2  y^2 -(\hat \mu_1+\hat \mu_2 + \hat\lambda) y +\hat\mu_2\label{def-b}\\
c(y) &:=& b_{-}(y) b_{+}(y)
\label{def-c}
\end{eqnarray}
with
\begin{equation}
b_{-}(y):=b(y)-2y\sqrt{\hat \lambda_1 \hat \mu_1},\quad
b_{+}(y):=b(y)+2y\sqrt{\hat\lambda_1 \hat\mu_1}.
\label{def-b+-}
\end{equation}
We have
\begin{equation}
\label{rootsb}
b_{-}(y)=\hat\lambda_2 (y-y_1)(y-y_4),\quad b_{+}(y)=\hat\lambda_2 (y-y_2)(y-y_3)
\end{equation}
with
\begin{eqnarray}
y_1 &=&\frac{\xi_1-\sqrt{\xi_1^2-4\hat\lambda_2\hat\mu_2}}{2\hat\lambda_2},\quad
y_2 =\frac{\xi_2-\sqrt{\xi_2^2-4\hat\lambda_2\hat\mu_2}}{2\hat\lambda_2}
\label{def:y1-y2}\\
y_3 &=&\frac{\xi_2+ \sqrt{\xi_2^2-4\hat\lambda_2\hat\mu_2}}{2\hat\lambda_2},\quad
y_4 =\frac{\xi_1+\sqrt{\xi_1^2-4\hat\lambda_2\hat\mu_2}}{2\hat\lambda_2}
\label{def:y3-y4}\\
\xi_1&=&\hat \mu_1 + \hat \mu_2 +\hat\lambda+2\sqrt{\hat\lambda_1\hat\mu_1}, \quad
\xi_2= \hat \mu_1 + \hat \mu_2 +\hat\lambda-2\sqrt{\hat\lambda_1\hat\mu_1}.
\label{def:xi1-xi2}
\end{eqnarray}
$y_1,\ldots,y_4$ are the branch points of $x(y)$ (since $c(y_i)=0$ for $i=1,\ldots,4$).
It is easily seen that (Hint: $y_2<1$ and $y_3>1$, both from Assumption {\bf A}))
\begin{equation}
\label{inq-y}
0<y_1<y_2<1< y_3 <y_4.
\end{equation}
\begin{remark}
\label{rem:circle}
The algebraic function $x(y)$ has two algebraic branches, denoted by $k(y)$ and $k^\sigma(y)$,
related via the relation $k(y)k^\sigma(y)=\hat\mu_1/\hat \lambda_1$. When $y\in (y_1,y_2)\cup (y_3,y_4)$
 $k(y)$ and $k^\sigma(y)$ are complex conjugate numbers $($since $c(y)<0$  for those values of $y$ $)$, with
$k(y_i)=k^\sigma(y_i)$ for $i=1,\ldots,4$.
In particular, $|k(y)|=\sqrt{k(y) k^\sigma(y)}=\sqrt{\hat\mu_1/\hat \lambda_1}$ for $y\in [y_1,y_2]\cup [y_3,y_4]$,
thereby showing that for $y\in [y_1,y_2]$ $($resp. $y\in [y_3,y_4]$ $)$ $k(y)$ and $k^\sigma(y)$ lie on the circle
centered in $0$ with radius $\sqrt{\hat\mu_1/\hat \lambda_1}$  .
\end{remark}

When $x$ is fixed similar results hold. We will denote by
\begin{equation}
\label{def:yx}
y(x)=\frac{-e(x)\pm\sqrt{d(x)}}{2 \hat\lambda_2 x}
\end{equation}
the algebraic function solution of $R(x,y)=0$ for $x$ fixed, where
$e(x):=\hat\lambda_1 x^2 -(\hat\mu_1+\hat\mu_2+\hat\lambda)x+\hat\mu_1$ and $d(x):= e_{-}(x)e_{+}(x)$, with
\[
e_{-}(x):= e(x)-2x\sqrt{\hat\lambda_2\hat\mu_2},\quad e_{+}(x):= e(x)+2x\sqrt{\hat\lambda_2\hat\mu_2}.
\]
We denote by $x_i$, $i=1,\ldots,4$ the four branch points of $y(x)$, namely, the zeros of $d(x)$;
they are obtained by interchanging indices
$1$ and $2$ in (\ref{def:y1-y2})-(\ref{def:xi1-xi2}).

We have
\begin{equation}
\label{def:e-+}
e_{-}(x)=\hat\lambda_1(x-x_1)(x-x_4),\quad e_{+}(x)=\hat\lambda_1(x-x_2)(x-x_3)
\end{equation}
where
\begin{equation}
0<x_1<x_2\leq 1 <x_3<x_4
\label{branchpoint-y(x)}
\end{equation}
with $x_2=1$ iff  $\hat\lambda_2=\hat\mu_2$.

The following results, found in \cite[Lemmas 2.2, 2.3, 3.1]{FI79}, hold :
\begin{proposition}
\label{prop:FayIas}
For $y$ fixed, the equation $R(x,y)=0$ has one root $x(y)=k(y)$ which is analytic in the whole complex plane cut $\mathbb{C}$
along $[y_1,y_2]$ and $[y_3,y_4]$.
Moreover\footnote{Apply Rouch\'e's theorem to $R(x,y)$ to get (a1), and the ``maximum modulus principal'' to the
analytic function $k(y)$ in $\mathbb{C}-[y_1,y_2]-[y_3,y_4]$ to get (b1). (c1) follows from Remark \ref{rem:circle}.}
 \begin{itemize}
\item[(a1)] $|k(y)|\leq 1$ if $|y|=1$. More precisely,
 $|k(y)|<1$ if  $|y|=1$ with $y\not=1$, and $k(1)=\min(1,\hat \mu_1/\hat \lambda_1)=1$ under Assumption {\bf A}.
\item[(b1)] $|k(y)|\leq \sqrt{\frac{\hat\mu_1}{\hat\lambda_1}}$ for all $y\in \mathbb{C}$;
\item[(c1)] when $y$ sweeps twice $[y_1,y_2]$, $k(y)$ describes a circle centered in $0$ with
radius $\sqrt{\frac{\hat\mu_1}{\hat\lambda_1}}$, so that
$|k(y)|=\sqrt{\frac{\hat\mu_1}{\hat\lambda_1}}$ for $y\in [y_1,y_2]$.
\end{itemize}
Similarly, for $x$ fixed, the equation $R(x,y)=0$ has one root $y(x)=h(x)$ which is analytic in
$\mathbb{C}-[x_1,x_2]-[x_3,x_4]$, and
 \begin{itemize}
\item[(a2)] $|h(x)|<1$ if $|x|=1$, $x\not=1$, and $h(1)=\min(1,\hat \mu_2/\hat \lambda_2)\leq 1$.
\item[(b2)] $|h(x)|\leq \sqrt{\frac{\hat\mu_2}{\hat\lambda_2}}$ for all $x\in \mathbb{C}$;
\item[(c2)] $|h(x))|=\sqrt{\frac{\hat\mu_2}{\hat\lambda_2}}$ if $x\in [x_1,x_2]$
\end{itemize}
Moreover,
 \begin{itemize}
\item[(d1)] $h(k(y))=y$ for $y\in [y_1,y_2]$ and (d2) $k(h(x))=x$ for $x\in [x_1,x_2]$.
\item[(d2)] $h(\sqrt{\hat\mu_1/\hat\lambda_1})=y_2$ and $h(-\sqrt{\hat\mu_1/\hat\lambda_1})=y_1$.
\item[(d3)]
$k(\sqrt{\hat\mu_2/\hat\lambda_2})=x_2$ and $k(-\sqrt{\hat\mu_2/\hat\lambda_2})=x_1$.
\end{itemize}
Last
 \begin{itemize}
\item[(e)] $|h(x)| \leq 1$ for $1\leq |x| \leq \sqrt{\frac{\hat\mu_1}{\hat\lambda_1}}$ (recall that $\hat\lambda_1<\hat\mu_1$).
\end{itemize}
\end{proposition}

\subsection{A boundary value problem and its solution}
\label{ssec:BVP}
We are now in a position to set a boundary value problem that is satisfied by the unknown function $H^{(0)}(x,0)$.

In the following, $C_a=\{z\in \mathbb{C}: |z|\leq a\}$ ($a>0$) denotes the circle centered in $0$ of radius $a$,
and $C_a^+=\{z\in \mathbb{C}: |z|< a\}$ denotes the interior of $C_a$.

We know that $R(k(y),y)=0$ by definition of $k(y)$. On the other hand,  $H^{(0)}(x,y)$ is well-defined
for all $(x,y)=(k(y), y)$ with $|y|=1$, since  (i) $H^{(0)}(x,y)$ is well-defined for $|x|\leq 1$, $|y|\leq 1$,
(ii) $k(y)$ is continuous for $|y|=1$ (from Proposition \ref{prop:FayIas} we know that $k(y)$ is analytic
in $\mathbb{C}-[y_1,y_2]$ and we know that $0<y_1<y_2<1$ so that $k(y)$ is continuous for $|y|=1$),
(iii) $|k(y)|\leq 1$ for $|y|=1$ (cf. Proposition \ref{prop:FayIas}-(a1)).
Therefore, the l.h.s. of (\ref{funct-eq}) must vanish
for all pairs $(k(y),y)$ such that $|y|=1$, which yields
\begin{equation}
\label{bvp0}
A(k(y),y)  H^{(0)}(k(y),0)=-B(k(y),y) H^{(0)}(0,y), \quad \forall |y|=1.
\end{equation}
The  r.h.s. of (\ref{bvp0}) is analytic for $|y|\leq 1$ with $y\not\in [y_1,y_2]$ and continuous for $|y|\leq 1$,
so that the r.h.s. of (\ref{bvp0}) can be analytically continued up to the interval $[y_1,y_2]$.

This gives
\begin{equation}
\label{bvp-1}
A(k(y),y)  H^{(0)}(k(y),0)=-B(k(y),y) H^{(0)}(0,y), \quad \forall y\in [y_1,y_2].
\end{equation}
It is shown in Lemma \ref{lem:zeroAB} that $B(k(y),y)\not=0$ for $y\in [y_1,y_2]$. We may therefore divide
both sides of (\ref{bvp-1}) by $B(k(y),y)$  to get
\begin{equation}
\label{bvp1}
\frac{A(k(y),y)}{B(k(y),y)} H^{(0)}(k(y),0)= - H^{(0)}(0,y), \quad \forall y\in [y_1,y_2].
\end{equation}
Take $y\in [y_1,y_2]$: we know by Proposition \ref{prop:FayIas}-(c1) that
$k(y)=x\in C_{\sqrt{\hat\mu_1/\hat\lambda_1}}$ so that $h(k(y))=y=h(x)\in [y_1,y_2]$
by Proposition \ref{prop:FayIas}-(d1). We may therefore rewrite (\ref{bvp1}) as
\begin{equation}
\label{bvp2}
\frac{A(x,h(x))}{B(x,h(x))} H^{(0)}(x,0) = -H^{(0)}(0,h(x)), \quad \forall x \in C_{\sqrt{\frac{\hat\mu_1}{\hat\lambda_1}}}.
\end{equation}
It is shown in Lemma \ref{lem:h} that $h(x)$ is analytic for $1<|x|<\sqrt{\hat\mu_1/\hat\lambda_1}$ and continuous
for $1\leq |x|\leq\sqrt{\hat\mu_1/\hat\lambda_1}$; furthermore
$|h(x)|\leq 1$ for $1\leq |x|\leq\sqrt{\hat\mu_1/\hat\lambda_1}$ by Proposition \ref{prop:FayIas}-(e). These two properties
imply that, $ H^{(0)}(0,h(x))$,  the r.h.s. of (\ref{bvp2}), is analytic for $1<|x|<\sqrt{\hat\mu_1/\hat\lambda_1}$
and continuous for $1\leq |x|\leq\sqrt{\hat\mu_1/\hat\lambda_1}$ , which in turn
implies that, $\frac{A(x,h(x))}{B(x,h(x))} H^{(0)}(x,0)$, the l.h.s. of (\ref{bvp2}), can be extended as a function that is analytic
for  $1<|x|<\sqrt{\hat\mu_1/\hat\lambda_1}$ and continuous for  $1\leq |x|\leq \sqrt{\hat\mu_1/\hat\lambda_1}$.

It is shown in Lemma \ref{lem:Azero} that $A(x,h(x))$ has exactly one zero in $(1,\sqrt{\hat\mu_1/\hat\lambda_1}]$, of multiplicity one,
given by
\begin{equation}
\label{def-x0}
x_0=\frac{-(\lambda +\mu_1-\mu)\lambda\mu_1+\sqrt{((\lambda +\mu_1-\mu)\lambda\mu_1)^2 + 4 \lambda \lambda_1(\lambda+\mu_1)\mu\mu_1^2}}
{2\lambda\lambda_1 (\lambda+\mu_1)},
\end{equation}
if $x_0\leq \sqrt{\hat\mu_1/\hat\lambda_1}$ and if $(\lambda +\mu_1)x_0/(\lambda x_0+\mu_1)\leq \sqrt{\hat\mu_2/\hat\lambda_2}$
and does not have any zero in $(1,\sqrt{\hat\mu_1/\hat\lambda_1}]$, otherwise.

Introduce
\begin{equation}
\label{def:U}
U(x):= \frac{A(x,h(x))}{B(x,h(x)) (x-x_0)^r} \quad \hbox{and} \quad \tilde H(x):=H^{(0)}(x,0) (x-x_0)^r,
\end{equation}
where $r\in \{0,1\}$ is defined by
\begin{equation}
\label{def:r}
r=\left\{  \begin{array}{ll}
1,  &if \,\mbox{$x_0 \le \displaystyle \sqrt{\hat\mu_1/\hat\lambda_1}$ and
$\displaystyle \frac{(\lambda +\mu_1)x_0}{\lambda x_0+\mu_1}\leq \sqrt{\hat\mu_2/\hat\lambda_2}$},\\
0,  &\mbox{otherwise.}
                     \end{array}
           \right.
\end{equation}

By construction
\begin{equation}
\label{eqU}
\frac{A(x,h(x))}{B(x,h(x))} H^{(0)}(x,0)= U(x) \tilde H(x).
\end{equation}
As noticed earlier the l.h.s. of (\ref{eqU})  is analytic
for  $1<|x|<\sqrt{\hat\mu_1/\hat\lambda_1}$ and continuous for  $1\leq |x|\leq \sqrt{\hat\mu_1/\hat\lambda_1}$. Since
by construction $U(x)$ does not vanish  in $(1,\sqrt{\hat\mu_1/\hat\lambda_1}]$ we conclude from (\ref{eqU})
that the function $\tilde H(x)$ that is initially analytic for $|x|<1$ and continuous for $|x|\leq 1$ can be extended
as a  function that is analytic for $|x|<\sqrt{\hat\mu_1/\hat\lambda_1}$ and continuous for
$|x|\leq \sqrt{\hat\mu_1/\hat\lambda_1}$.

In summary, we have shown that the real part
\begin{equation}
\label{bvp4}
\Re\left(i \,U(x) \tilde H(x)\right)=0, \quad \forall x \in C_{\sqrt{\hat\mu_1/\hat\lambda_1}},
\end{equation}
where $\tilde H(x)$ is analytic in $C^+_{\sqrt{\hat\mu_1/\hat\lambda_1}}$ and continuous
in $C^+_{\sqrt{\mu_1/\hat\lambda_1}}\cup C_{\sqrt{\hat\mu_1/\hat\lambda_1}}$, and where
$U(x)$ does not vanish on $C_{\sqrt{\hat\mu_1/\hat\lambda_1}}$.
 This defines  a Riemann-Hilbert boundary value problem on the circle $C_{\sqrt{\hat\mu_1/\hat\lambda_1}}$, whose solution
 is given below.

 Define
\begin{equation}
\label{def-index}
\chi:=-\frac{1}{\pi}[\hbox{arg } U(x)]_{x\in C_{\sqrt{\hat\mu_1/\hat\lambda_1}}}
\end{equation}
the so-called index of the Riemann-Hilbert problem, where $[\hbox{arg}\,\alpha(z)]_{z\in C}$ denotes the variation of the
argument of the function $\alpha(z)$ when $z$ moves on a closed curved $C$ in the positive direction, provided
that $\alpha(z)\not=0$ for $z\in C$).

The Riemann-Hilbert problem has $\chi+1$ independent solutions \cite[p. 104]{Musk}. It is shown  in
Lemma \ref{lem:index} that, as expected, $\chi=0$ under conditions (\ref{stability}), thereby showing that the solution
of the  Riemann-Hilbert problem (\ref{bvp4}) is unique under  conditions (\ref{stability}) which will in turn
imply that (\ref{stability}) are sufficient stability conditions for the queueing system at hand.

With $\chi=0$ the solution of the Riemann-Hilbert problem is
\begin{equation}
H^{(0)}(x,0)=D (x-x_0)^{-r} \exp\left(\frac{1}{2\pi i}\int_{|z|=\sqrt{\hat\mu_1/\hat\lambda_1}} \frac{\log(J(z))}{z-x}dz \right), \quad
\forall\, |x|<\sqrt{\hat\mu_1/\hat\lambda_1},
\label{sol-bvp}
\end{equation}
where $D$ is  a constant (to be determined) and (with $\overline{z}$ the complex conjugate of $z\in \mathbb{C}$)
\[
J(z)=-\frac{\overline{i U(z)}}{i U(z)}.
\]
We are left with calculating the constant $D$ in (\ref{sol-bvp}). Setting $x=1$ in (\ref{sol-bvp}) gives
\begin{equation}
\label{eq:D}
D=(1-x_0)^{r} \left(1-\frac{\lambda}{\mu}\left(1+\frac{\lambda_2}{\mu_2}\right)\right)
\exp\left(-\frac{1}{2\pi i}\int_{|z|=\sqrt{\hat\mu_1/\hat\lambda_1}} \frac{\log(J(z))}{z-1}dz \right)
\end{equation}
by using the value of $H^{(0)}(1,0)$ found in (\ref{eq:109-2}). We  may therefore rewrite (\ref{sol-bvp}) as
\begin{equation}
\label{final-bvp}
H^{(0)}(x,0)=\left(\frac{1-x_0}{x-x_0}\right)^r \left(1-\frac{\lambda}{\mu}\left(1+\frac{\lambda_2}{\mu_2}\right)\right)
\exp\left(\frac{1}{2\pi i}\int_{|z|=\sqrt{\hat\mu_1/\hat\lambda_1}} \frac{\log(J(z))(x-1)}{(z-x)(z-1)}dz \right)
\end{equation}
for all $|x|<\sqrt{\hat\mu_1/\hat\lambda_1}$.

We also need to calculate the other boundary function $H^{(0)}(0,y)$ for $|y|\leq 1$.
For $|y|=1$,  $H^{(0)}(0,y)$ is given in (\ref{bvp0}). For $|y|<1$,
$H^{(0)}(0,y)$ is obtained  from  (\ref{bvp0}) and  Cauchy's formula, which gives
\begin{equation}\label{eq:H00y}
H^{(0)}(0,y) =  \frac{1}{2\pi i} \int_{|t|=1} \frac{V(t)}{t-y} dt, \quad |y|<1,
\end{equation}
where
\begin{equation}
\label{dev:Vt}
V(t) := - \frac{A(k(t),t)}{B(k(t),t)} H^{(0)}(k(t),0), \quad |t|=1,
\end{equation}
does not vanish for all $|t|=1$, as shown in  Lemma~\ref{lem:Bzero}.

Introducing (\ref{final-bvp}) and (\ref{eq:H00y}) into (\ref{eq:Mform}) uniquely determines
the joint generating functions
$H^{(0)}(x,y)$ and $H^{(1)}(x,y)$ for $|x|\leq 1$, $|y|\leq 1$ which shows, as announced,
that conditions (\ref{stability}) are also sufficient for the system to be stable.

\section{Performance measures}

Later on in this section we shall need the derivatives $\frac{d}{dx}H^{(0)}(x,0)|_{x=1}$ and
$\frac{d}{dy}H^{(0)}(0,y)|_{y=1}$.

Differentiating  (\ref{final-bvp}) w.r.t $x$ gives
\begin{eqnarray}
\frac{d}{dx}  H^{(0)}(x,0) &=&\left( \frac{1-x_0}{x-x_0} \right)^{r}
\left(   1-\frac{\lambda}{\mu} \left(1+\frac{\lambda_2}{\mu_2}\right) \right)\nonumber\\
&\times&
\exp\left(\frac{1}{2\pi i}\int_{|z|=\sqrt{\hat\mu_1/\hat\lambda_1}} \frac{\log(J(z))(x-1)}{(z-x)(z-1)}dz \right)
\nonumber\\
&\times &\left(\frac{-r}{x-x_0} +\frac{1}{2\pi i} \int_{|z|=\sqrt{\hat\mu_1/\hat\lambda_1}}\frac{\log(J(z)}{(z-x)^2}dz\right)\nonumber\\
&=&
H^{0}(x,0)\left( \frac{-r}{x-x_0} +\frac{1}{2\pi i} \int_{|z|=\sqrt{\hat\mu_1/\hat\lambda_1}}\frac{\log(J(z)}{(z-x)^2}dz\right).
\label{derivativeHx0}
\end{eqnarray}
Letting $x=1$ in (\ref{derivativeHx0}) and using (\ref{eq:109-2}) yields
\begin{equation}
\label{derivativeH10}
\frac{d}{dx}  H^{(0)}(x,0) |_{x=1}=\left(   1-\frac{\lambda}{\mu} \left(1+\frac{\lambda_2}{\mu_2}\right) \right)
\left( \frac{r}{x_0-1} +\frac{1}{2\pi i} \int_{|z|=\sqrt{\hat\mu_1/\hat\lambda_1}}\frac{\log(J(z)}{(z-1)^2}dz\right).
\end{equation}
The derivative $\frac{d}{dy}H^{(0)}(0,y)|_{y=1}$ is obtained from (\ref{bvp0}). By Lemma~\ref{lem:Bzero}, we have
\begin{eqnarray}
\frac{d}{dy}H^{(0)}(0,y)|_{y=1} &=&
-\lim_{y \to 1} \frac{A(k(y),y)}{B(k(y),y)}  \ \frac{d}{dx}H^{(0)}(x,0)|_{x=1} \  k'(1) \nonumber\\
&&
-\lim_{y \to 1} \frac{d}{dy} \frac{A(k(y),y)}{B(k(y),y)} \ H^{(0)}(1,0),
\label{derivativeH01}
\end{eqnarray}
where $\frac{d}{dx}H^{(0)}(x,0)|_{x=1}$ and $H^{(0)}(1,0)$ are given in (\ref{derivativeH10}) and (\ref{eq:109-2}), respectively.
The limits in the above expression can be calculated by L'H\^opital's rule. Lengthy but easy algebra gives
$$
\lim_{y \to 1} \frac{A(k(y),y)}{B(k(y),y)} =
\frac{(\lambda_2-\mu+\lambda_1 k'(1))\mu_2}{(\lambda_2+(\lambda_1-\mu)k'(1))\mu_1}
$$
and
$$
\lim_{y \to 1} \frac{d}{dy} \frac{A(k(y),y)}{B(k(y),y)} =
$$
$$
-\frac{(-\lambda_2+(-\lambda_1+\mu)k'(1)+(\lambda_2-\mu)k'(1)^2+\lambda_1k'(1)^3+(\mu-\lambda_1-\lambda_2)k''(1))\mu\mu_2}
{(\lambda_2+(\lambda_1-\mu)k'(1))\mu_1},
$$
where
$$
k'(1)=\frac{\hat\lambda_2- \hat\mu_2}{\hat\mu_1-\hat\lambda_1},
$$
and
$$
k''(1)=2\frac{(\hat\mu_1+\hat\mu_2-2(\hat\lambda_1+\hat\lambda_2))\hat\mu_1\hat\mu_2
+\hat\lambda_1^2\hat\mu_2+\hat\lambda_2^2\hat\mu_1}{(\hat\mu_1-\hat\lambda_1)^3}.
$$
We are now in a position to calculate some important performance measures.

By setting $x=0$ in equation (\ref{final-bvp}), we
immediately obtain the probability of empty system
\begin{eqnarray}
\lefteqn{P(Q_1=0,Q_2=0,L=0) =
\left(\frac{x_0-1}{x_0}\right)^r \left(1-\frac{\lambda}{\mu}\left(1+\frac{\lambda_2}{\mu_2}\right)\right)}\nonumber\\
&&\times
\exp\left(\frac{1}{2\pi i}\int_{|z|=\sqrt{\hat\mu_1/\hat\lambda_1}} \frac{\log(J(z))}{z(1-z)}dz \right)
\label{empty}
\end{eqnarray}

Next, we calculate the expected orbit queue lengths. For the first queue, we have
\begin{equation}
\label{eq:Q1}
E[Q_1] =
\sum_{m=1}^\infty m \left( \sum_{n=0}^\infty P_{mn}(0) + \sum_{n=0}^\infty P_{mn}(1)\right)
= \frac{d}{dx} H^{(0)} (x,1)|_{x=1} + \frac{d}{dx} H^{(1)} (x,1)|_{x=1}.
\end{equation}
Thus, we need to calculate $\frac{d}{dx} H^{(0)}(x,1)|_{x=1}$ and $\frac{d}{dx} H^{(1)}(x,1)|_{x=1}$.
From  (\ref{funct-eq}) we have
\begin{equation}
\label{H0xy}
H^{(0)}(x,y) = \frac{A(x,y)}{R(x,y)} H^{(0)}(x,0) + \frac{B(x,y)}{R(x,y)}H^{(0)}(0,y).
\end{equation}
Using (\ref{def-K})-(\ref{def-B}) and setting $y=1$ in (\ref{H0xy}), yields
$$
H^{(0)}(x,1) = \frac{\lambda_1 \mu_2 x}{\alpha\lambda_1 x - \mu\mu_1} H^{(0)}(x,0)
+ \frac{(\lambda_1 x - \mu)\mu_1}{\alpha\lambda_1 x - \mu\mu_1}H^{(0)}(0,1).
$$
Next, by differentiating the above relation with respect to $x$ we get
$$
\frac{d}{dx} H^{(0)}(x,1) =
-\frac{\lambda_1\mu_2\mu\mu_1}{(\alpha\lambda_1 x - \mu\mu_1)^2} H^{(0)}(x,0)
+\frac{\lambda_1\mu_2 x}{\alpha\lambda_1 x - \mu\mu_1} \frac{d}{dx} H^{(0)}(x,0)
$$
$$
+\frac{\lambda_1\mu_1\mu(\alpha-\mu_1)}{(\alpha\lambda_1 x - \mu\mu_1)^2} H^{(0)}(0,1).
$$
Setting $x=1$ in the above, yields
\begin{eqnarray}
\frac{d}{dx} H^{(0)}(x,1)|_{x=1} &=&
\frac{\lambda_1\mu_1\mu}{(\mu\mu_1-\alpha\lambda_1)^2} \left((\alpha-\mu_1)H^{(0)}(0,1)
-\mu_2 H^{(0)}(1,0)\right) \nonumber\\
&&-\frac{\lambda_1\mu_2}{\mu\mu_1-\alpha\lambda_1} \frac{d}{dx} H^{(0)}(x,0)|_{x=1},
\label{dH011}
\end{eqnarray}
where $H^{(0)}(0,1)$, $H^{(0)}(1,0)$ and $dH^{(0)}(x,0)/dx|_{x=1}$
are given in (\ref{eq:109-1}), (\ref{eq:109-2}) and (\ref{derivativeH10}), respectively.

It remains to find $d H^{(1)}(x,1)/dx|_{x=1}$.
Differentiating (\ref{eq1}) with respect to $x$ and setting $x=y=1$ gives
\begin{eqnarray}
\frac{d}{dx} H^{(1)}(x,1)|_{x=1} &=& \frac{\alpha}{\mu} \frac{d}{dx} H^{(0)}(x,1)|_{x=1}
-\frac{\mu_2}{\mu} \frac{d}{dx} H^{(0)}(x,0)|_{x=1}\nonumber\\
&=&\frac{\alpha\lambda_1\mu_1}{(\mu\mu_1-\alpha\lambda_1)^2} \left((\alpha-\mu_1)H^{(0)}(0,1)
-\mu_2 H^{(0)}(1,0)\right)\nonumber\\
&&
-\frac{\mu_1\mu_2}{\mu\mu_1-\alpha\lambda_1} \frac{d}{dx} H^{(0)}(x,0)|_{x=1},
\label{dH111}
\end{eqnarray}
by using (\ref{dH011}).

By combining (\ref{eq:Q1}), (\ref{dH011}) and (\ref{dH111}) we finally obtain
\begin{eqnarray}
E[Q_1] &=&
\frac{(\alpha+\mu)\lambda_1\mu_1}{(\mu\mu_1-\alpha\lambda_1)^2} \left((\alpha-\mu_1)H^{(0)}(0,1)
-\mu_2 H^{(0)}(1,0)\right)\nonumber\\
&&-
\frac{\mu_2(\lambda_1+\mu_1)}{\mu\mu_1-\alpha\lambda_1} \frac{d}{dx} H^{(0)}(x,0)|_{x=1},
\label{EQ1subs}
\end{eqnarray}
where $H^{(0)}(0,1)$, $H^{(0)}(1,0)$ and $dH^{(0)}(x,0)/dx|_{x=1}$
are given in (\ref{eq:109-1}), (\ref{eq:109-2}) and (\ref{derivativeH10}), respectively.

Similarly, the expected queue length for the second orbit is given by
\begin{eqnarray}
E[Q_2] &=& \frac{d}{dy} H^{(0)} (1,y)|_{y=1} + \frac{d}{dy} H^{(1)} (1,y)|_{y=1}\nonumber\\
&=&
\frac{(\alpha+\mu)\lambda_2\mu_2}{(\mu\mu_2-\alpha\lambda_2)^2} \left((\alpha-\mu_2)H^{(0)}(1,0)
-\mu_1 H^{(0)}(0,1)\right)\nonumber\\
&&
-\frac{\mu_1(\lambda_2+\mu_2)}{\mu\mu_2-\alpha\lambda_2} \frac{d}{dy} H^{(0)}(0,y)|_{y=1},
\label{EQ2subs}
\end{eqnarray}
where $dH^{(0)}(0,y)/dy|_{y=1}$ is given in (\ref{derivativeH01}).

Finally, we recall that (see (\ref{eq:PL1}))
$$
E[L] = P(L=1) = \frac{\lambda}{\mu}.
$$

\section{Numerical examples}

\begin{figure}[t]
\centering
\includegraphics[width=12cm]{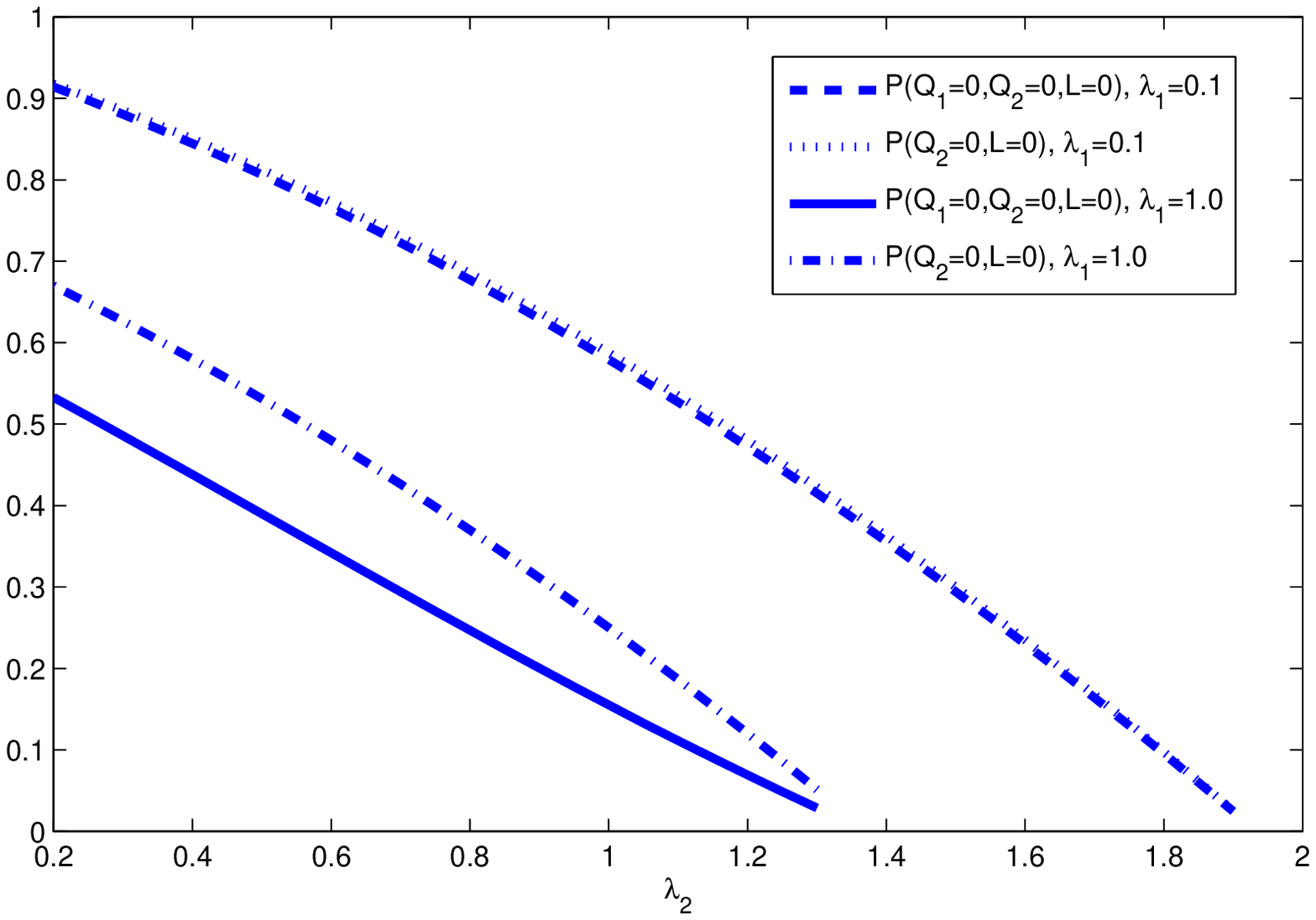}
\caption{Probability of an empty system ($\mu=4$, $\mu_1=\mu_2=2$).}
\label{fig:P000}
\end{figure}

To obtain more insights into the performance of the system, let us consider numerical examples. First,
we set $\mu_1=\mu_2=2$, $\mu=4$, $\lambda_1=0.1$ and
vary $\lambda_2$ in the interval $[0.2;1.9]$. In Figure~\ref{fig:P000} we plot the probability of an empty
system $P(Q_1=0,Q_2=0,L=0)$ calculated by (\ref{empty}) as a function of $\lambda_2$.
We also plot $H^{(0)}(1,0)$, see formula (\ref{eq:109-2}), which corresponds, if $\lambda_1$ is small,
to the probability of empty system with one type of jobs and a single orbit queue.
Now if we change the value of $\lambda_1$ from 0.1 to 1.0, we observe that the value of $P(Q_1=0,Q_2=0,L=0)$
deviates significantly from $H^{(0)}(1,0)$.

\begin{figure}[t]
\centering
\includegraphics[width=12cm]{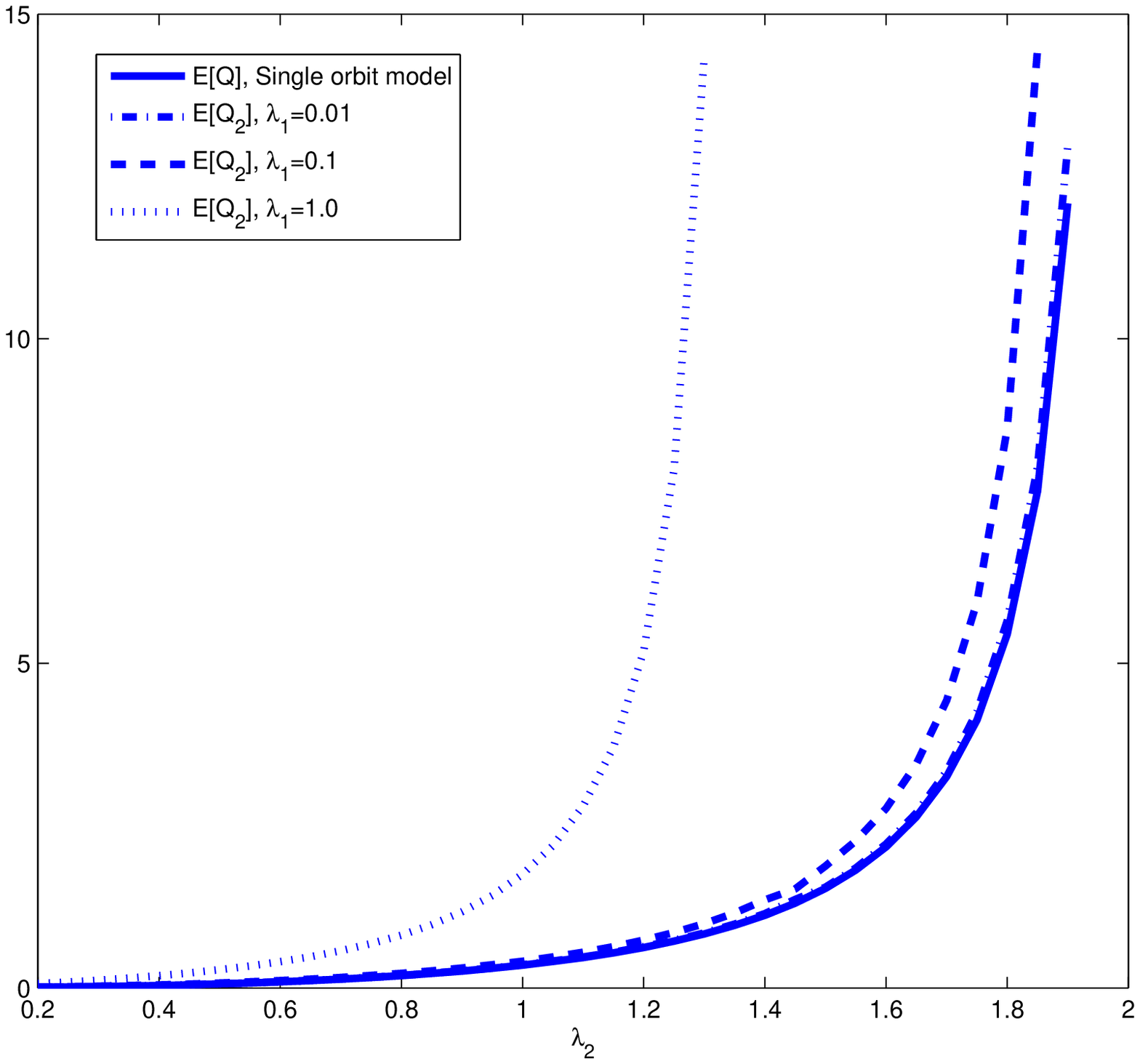}
\caption{The expected orbit queue size, $E[Q_2]$ ($\mu=4$, $\mu_1=\mu_2=2$).}
\label{fig:EQSingleOrbit}
\end{figure}

Keeping $\mu_1=\mu_2=2$, $\mu=4$, in Figure~\ref{fig:EQSingleOrbit} we plot the expected queue length
of the second orbit $E[Q_2]$ calculated by (\ref{EQ2subs}) as a function of $\lambda_2$ for $\lambda_1=0.01;0.1;1.0$.
We also plot the expected queue length of the orbit queue for the single orbit retrial
system \cite{AY08}, which is given by
$$
E[Q] = \frac{\lambda_2^2(\lambda_2+\mu+\mu_2)}{\mu(\mu \mu_2 - \lambda_2^2 - \lambda_2 \mu_2)}.
$$
Again, as expected, when $\lambda_1$ goes to zero, $E[Q_2]$ approaches the expected queue length of the orbit queue
in the single orbit retrial system.

\begin{figure}[t]
\centering
\includegraphics[width=12cm]{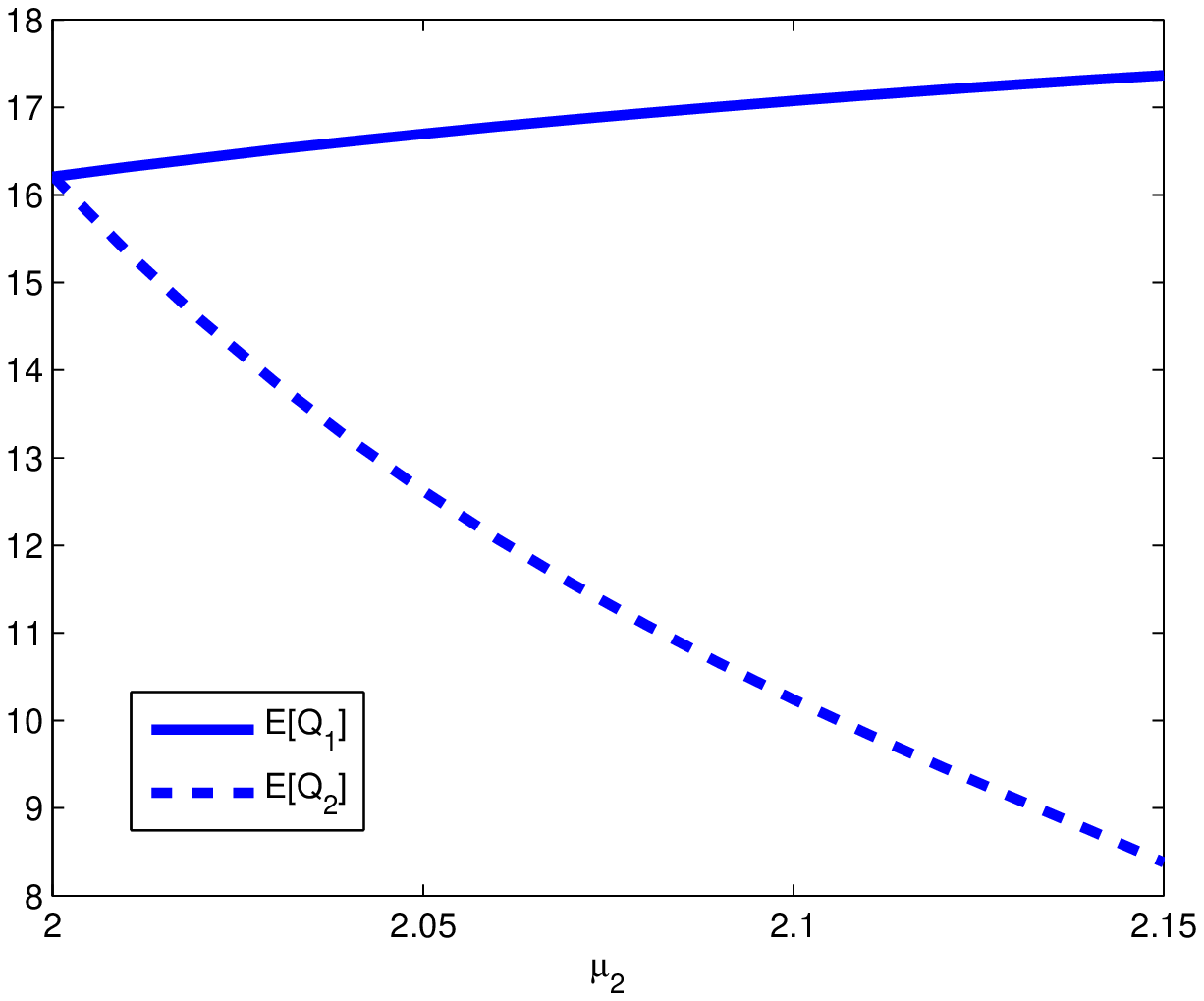}
\caption{The expected queue lengths of the orbit queues as functions of $\mu_2$
($\lambda_1=\lambda_2=1.2$, $\mu=4$, $\mu_1=2$).}
\label{fig:EQ1EQ2mu2}
\end{figure}

\begin{figure}[t]
\centering
\includegraphics[width=12cm]{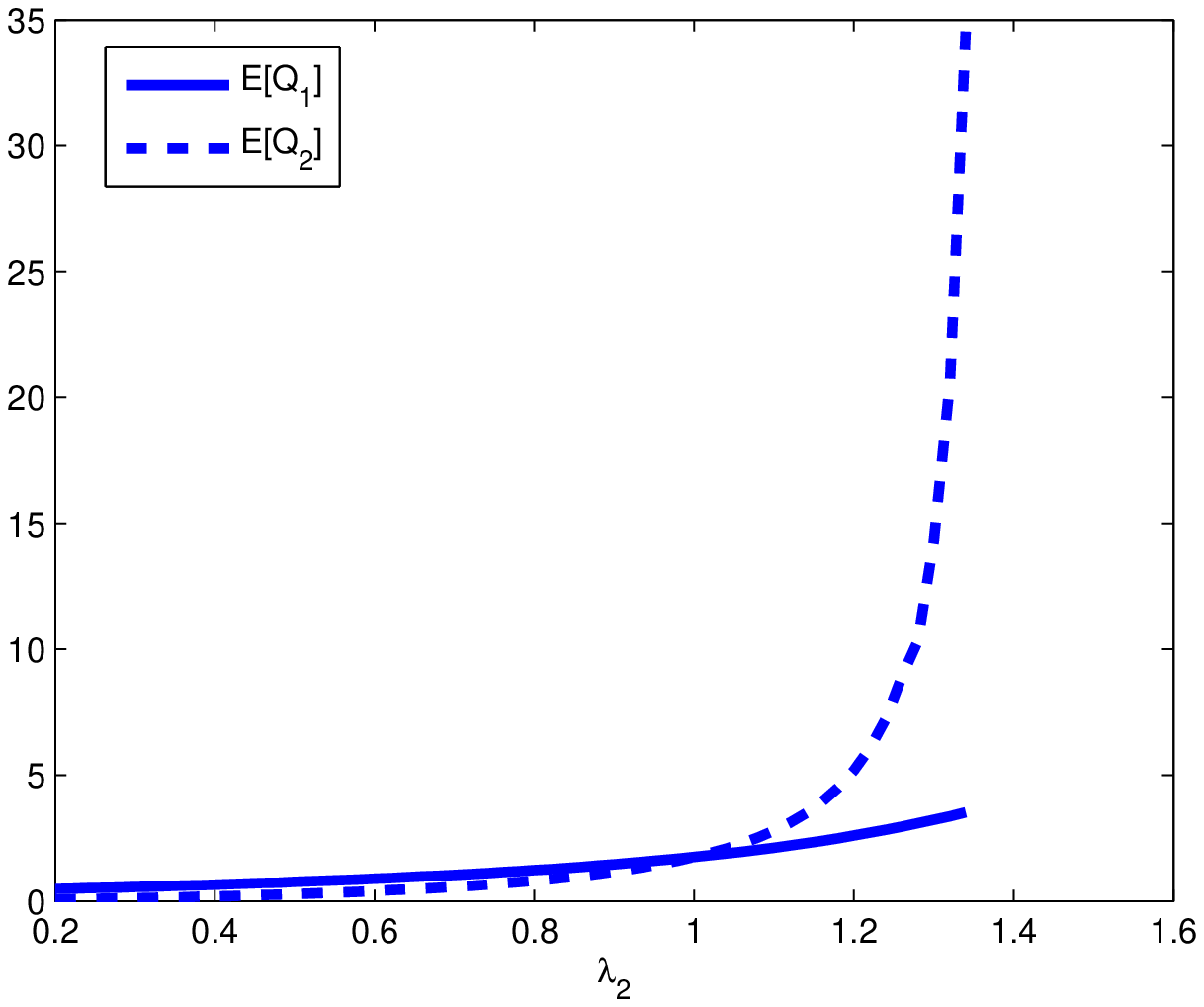}
\caption{The expected queue lengths of the orbit queues as functions of $\lambda_2$
($\lambda_1=1$, $\mu_1=\mu_2=2$, $\mu=4$).}
\label{fig:EQ1EQ2}
\end{figure}

Next, we investigate how the retrial rates affect the system performance. Let us fix $\lambda_1=\lambda_2=1.2$,
$\mu=4$, $\mu_1=2$ and we vary $\mu_2$ in the interval $[2.0;2.15]$. With such parameter setting,
the system is not too far from the stability boundary. We plot in Figure~\ref{fig:EQ1EQ2mu2}
the expected lengths of the orbit queues, $E[Q_1]$ and $E[Q_2]$, as functions of $\mu_2$.
We can see that if the jobs of type~2 retry at a bit faster rate than the jobs of type~1,
they can gain significantly in terms of the waiting time. Specifically, an increase of less than
10\% of the retrial rate of jobs of type~2 helps them to reduce the expected orbit queue length
by 50\%. Clearly, if there is no cost for retrials, it is beneficial for the jobs to increase their
retrial rate. However, there are good reasons to keep the control of the retrial rates in the hand
of the system administrator and not to set them too high. As was just mentioned, the first reason
is the possible cost for retrials. The second reason is the creation of incentives to respect the contract.
To illustrate this point, we fix $\lambda_1=1$, $\mu_1=\mu_2=2$, $\mu=4$, and vary $\lambda_2$
in the interval $[0.2;1.34]$. In Figure~\ref{fig:EQ1EQ2}, we plot the expected queue lengths of the orbit queues.
We see that if the jobs of type~2 increase their input rate beyond their fair share, they will be severely
penalized in terms of the expected delay, whereas the increase of the input rate of jobs of type~2 does
not inflict any significant damage to the jobs of type~1.

\section*{Acknowledgement}

We would like to thank Efrat Perel for helping us to draw the figure of the transition-rate diagram.

\appendix
\section{Appendix}

\begin{lemma}
\label{lem:zeroAB}
Under conditions (\ref{stability}), (i) $A(k(y),y)\not=0$ and (ii) $B(k(y),y)\not=0$ for\\ $y \in [y_1,y_2]$.

Equivalenty, (iii) $A(x,h(x))\not=0$ and (iv) $B(x,h(x))\not=0$ for $x\in C_{\sqrt{\hat \mu_1/\hat\lambda_1}}$.

\end{lemma}
{\bf Proof.}
From (\ref{def-B}) and (\ref{R-B}) we see that
$R(x,y)$ and $B(x,y)$ vanish simultaneously if and only if
\begin{eqnarray*}
(1-x)(\lambda_1 x -\mu)+\lambda_2(1-y)x&=&0\\
\lambda (1-x)y +\mu_2(y-x)&=&0.
\end{eqnarray*}
The second equation gives $x=(\lambda+\mu_2)y/((\lambda y +\mu_2))$. Plugging this value of $x$ into the
first equation yields (Hint: use $\lambda=\lambda_1+\lambda_2$)
\[
P_1(y):=(1-y) Q_1(y)
\]
with $Q_1(y):=\lambda \lambda_2 (\lambda+\mu_2)y^2 +(\lambda+\mu_2-\mu)\lambda \mu_2 y -\mu \mu_2^2=0$.

From $\lim_{y\to \pm \infty}Q_1(y)=+\infty$ and $Q_1(0)=-\mu \mu_2^2$ we conclude that
the polynomial $Q_1(y)$ has two real roots, $y_{-}<0<y_+$ and that $Q_1(y)<0$ for $0\leq y<y_+$.
Since
\begin{equation}
\label{inqQ}
Q_1(1)=\left(\frac{\lambda+\mu_2}{\mu \mu_2}\right)
\left(\frac{\lambda}{\mu}\left(1+\frac{\lambda_2}{\mu_2}\right)-1\right)<0,
\end{equation}
where the latter inequality holds under conditions (\ref{stability}),
we conclude that $Q_1(y)<0$ for $y\in [0,1]$, which in turn implies that $P_1(y)<0$ for  $y\in [0,1)$.
The latter completes the proof of (ii) since $[y_1,y_2]\subset [0,1)$ (see (\ref{inq-y})).

The proof of (i) is the same as the proof of (ii) up to interchanging incides $1$ and $2$.

Eqns (iii) and (iv) both follow from the fact that $k([y_1,y_2])
=C_{\sqrt{\hat\mu_1/\hat\lambda_1}}$ (cf. Proposition \ref{prop:FayIas}-(11))
and the relation $h(k(y))=y$ for $y\in [y_1,y_2]$
(cf. Proposition \ref{prop:FayIas}-(d1)).
\hfill\done

\begin{lemma}
\label{lem:h}
Under condition {\bf A}, $h(x)$ is analytic for $1<|x|<\sqrt{\hat\mu_1/\hat\lambda_1}$ and continuous
for  $1\leq|x|\leq\sqrt{\hat\mu_1/\hat\lambda_1}$
\end{lemma}

{\bf Proof.} We already know by Proposition \ref{prop:FayIas} that $h(x)$  is analytic for $x\in\mathbb{C}-[x_1,x_2]-[x_3,x_4]$
where $x_2\leq 1<x_3$.  It is therefore enough to show that  $\sqrt{\hat\mu_1/\hat\lambda_1}<x_3$ or, equivalently from
(\ref{def:e-+}) that $e_{+}\left(\sqrt{\hat\mu_1/\hat\lambda_1}\right)<0$.
Easy algebra shows that $e_+\left(\sqrt{\hat\mu_1/\hat\lambda_1}\right)
=-\sqrt{\hat\mu_1/\hat \lambda_1} \left( \left(\sqrt{\hat\lambda_1}-\sqrt{\hat\mu_1}\right)^2 +
\left(\sqrt{\hat\lambda_2}+\sqrt{\hat\mu_2}\right)^2\right)<0$, which concludes the proof.
\hfill\done

\begin{lemma}
\label{lem:Azero}
Assume that conditions (\ref{stability}) hold. Define
\[
x_0:=\frac{-(\lambda +\mu_1-\mu)\lambda\mu_1+\sqrt{((\lambda +\mu_1-\mu)\lambda\mu_1)^2 + 4 \lambda \lambda_1(\lambda+\mu_1)\mu\mu_1^2}}
{2\lambda\lambda_1 (\lambda+\mu_1)} >1
\]
If $x_0\leq \sqrt{\hat\mu_1/\hat\lambda_1}$ and if $(\lambda+\mu_1)x_0/(\lambda x_0+\mu_1)\leq
\sqrt{\hat\mu_2/\hat\lambda_2}$ then $A(x,h(x))$ has exactly one zero $x=x_0$ in $\left(1,\sqrt{\hat\mu_1/\hat\lambda_1}\right]$
and this zero has multiplicity one.
Otherwise  $A(x,h(x))$ has no zero in $\left(1,\sqrt{\hat\mu_1/\hat\lambda_1}\right]$.
\end{lemma}

{\bf Proof.}
From (\ref{def-A}) and (\ref{R-A}) we see that
$R(x,y)$ and $A(x,y)$ vanish simultaneously if and only if
\begin{eqnarray*}
(1-y)(\lambda_2 y -\mu)+\lambda_1(1-x)y&=&0\\
\lambda (1-y)x +\mu_1(x-y)&=&0.
\end{eqnarray*}
The second equation gives
\begin{equation}
\label{def-y}
y=\frac{(\lambda+\mu_1)x}{\lambda x +\mu_1}.
\end{equation}
Plugging this value of $y$ into the first equation yields
\[
P_2(x):=\frac{1-x}{(\lambda x +\mu_1)^2}\,Q_2(x)
\]
with $Q_2(x):=\lambda\lambda_1 (\lambda+\mu_1) x^2 +(\lambda+\mu_1-\mu) \lambda\mu_1 x -\mu\mu_1^2$.

The polynomial $Q_2(x)$ has exactly one positive zero given by $x_0$. From the inequality
\[
Q_2(1)=\mu \mu_1 (\lambda +\mu_1) \left(\frac{\lambda}{\mu}+ \frac{\lambda\lambda_1}{\mu\mu_1}-1\right) < 0
\]
which holds from (\ref{stability}), together with $\lim_{x\to\pm \infty}Q_2(x)=+\infty$ and $Q_2(0)<0$,
we conclude that $1<x_0$.

This shows that
\begin{itemize}
\item[-]
If $x_0> \sqrt{\hat\mu_1/\hat\lambda_1}$ then $A(x,h(x))$ has no zero in $(1, \sqrt{\hat\mu_1/\hat\lambda_1}]$;
\item[-]
Assume that $x_0 \leq  \sqrt{\hat\mu_1/\hat\lambda_1}$.  $A(x,h(x))$ as a unique zero in  $(1, \sqrt{\hat\mu_1/\hat\lambda_1}]$,
given by $x=x_0$ {\em provided} that (see (\ref{def-y})) $h(x_0)=(\lambda+\mu_1)x_0/(\lambda x_0+\mu_1)\leq \sqrt{\hat\mu_2/\hat\lambda_2}$
since we know from Proposition \ref{prop:FayIas}-(b2) that the branch $h(x)$
is such that $|h(x)|\leq  \sqrt{\hat\mu_2/\hat\lambda_2}$ for all $x\in \mathbb{C}$; if
$(\lambda+\mu_1)x_0/(\lambda x_0+\mu_1)>\sqrt{\hat\mu_2/\hat\lambda_2}$ then $A(x,h(x))$ does not vanish in $(1, \sqrt{\hat\mu_1/\hat\lambda_1}]$.
\end{itemize}
We are left with proving that when  $A(x,h(x))$ vanishes at $x=x_0$ then this zero has multiplicity one.
From now on we assume that $A(x_0,h(x_0))=0$.

From the definition of $h(x)$ and (\ref{R-A}) we get
\[
0=R(x,h(x))=\frac{\alpha}{\mu_2} A(x,h(x))+\mu[\lambda (1-h(x))x +\mu_1(x-h(x))].
\]
Differentiating this equation w.r.t. $x$ gives
\begin{equation}
\label{eq10}
0=\frac{\alpha}{\mu_2} \frac{dA(x,h(x))}{dx} +\mu [-\lambda h'(x)x +\lambda(1-h(x))+\mu_1(1-h'(x))].
\end{equation}
Assume that  $dA(x,h(x))/dx=0$ at point $x=x_0$, namely, assume that $A(x,h(x))$ has a zero of multiplicity
at least two at $x=x_0$. From (\ref{eq10}) this implies
\[
-\lambda h'(x_0)x_0 +\lambda(1-h(x_0))+\mu_1(1-h'(x_0)=0
\]
that is
\begin{equation}
\label{eq11}
h'(x_0)=\mu_1 \frac{\lambda+\mu_1}{(\lambda x_0+\mu_1)^2}
\end{equation}
with $h(x_0)=(\lambda+\mu_1)x_0/(\lambda x_0+\mu_1)$ (see (\ref{def-y}).

On the other hand, letting $(x,y)=(x,h(x))$ in (\ref{def-A}) yields
\begin{equation}
\label{eq4}
A(x,h(x))= ((1-h(x))(\lambda_2 h(x)-\mu)+\lambda_1 (1-x)h(x))\mu_2 x.
\end{equation}
Differentiating $A(x,h(x)$ wrt $x$ in (\ref{eq4}) and letting $x=x_0$ gives
\begin{eqnarray*}
\lefteqn{\frac{dA(x,h(x))}{dx}|_{x=x_0}=}\\
&& [ -h'(x_0) (\lambda h(x_0)-\mu)+\lambda_2(1-h(x_0))h'(x_0)-\lambda_1 h(x_0)+\lambda_1(1-x_0)h'(x_0)]\mu_2 x_0
\\
&& + \frac{\mu_2}{x_0}A(x_0,h(x_0))\\
&=& [h'(x_0)(-2\lambda_2 h(x_0)+\lambda_2+\mu +\lambda_1(1-x_0))-\lambda_1 h(x_0)]\mu_2 x_0
\\
&& + \frac{\mu_2}{x_0}A(x_0,h(x_0))\\
&=&
[h'(x_0)(-2\lambda_2 h(x_0)+\lambda_2+\mu +\lambda_1(1-x_0))-\lambda_1 h(x_0)]\mu_2 x_0
\end{eqnarray*}
since $A(x_0,h(x_0))=0$. Therefore,  $dA(x,h(x))/dx=0$ at point $x=x_0$ iff (note that $x_0\not=0$)
\[
h'(x_0)(-2\lambda_2 h(x_0)+\lambda_2+\mu +\lambda_1(1-x_0))-\lambda_1 h(x_0)=0.
\]
Since $-2\lambda_2 h(x_0)+\lambda_2+\mu +\lambda_1(1-x_0)<0$ because $x_0>1$, we get
\[
h'(x_0)=\frac{\lambda_1 h(x_0)}{-2\lambda_2 h(x_0)+\lambda_2+\mu +\lambda_1(1-x_0)}
\]
with (see (\ref{def-y}) $h(x_0)=(\lambda+\mu_1)x_0/(\lambda x_0+\mu_1)$, so that  $h'(x_0)<0$.
However, $h'(x_0)>0$ in (\ref{eq11}). This yields a contradiction,
thereby implying that  $dA(x,h(x))/dx$ does not vanish at point $x=x_0$ when $A(x,h(x))$ does
or, equivalently, that $x_0$ is a zero of multiplicity one.
\hfill\done

\begin{lemma}
\label{lem:index}
Under conditions (\ref{stability}) and Assumption {\bf A} the index $\chi$
of the Riemann-Hilbert problem (the index is defined in (\ref{def-index})) is equal to zero.
\end{lemma}

{\bf Proof.}

Recall the definition of $U(x)$ in (\ref{def:U}). First, by studying
$U(\sqrt{\hat\mu_1/\hat\lambda_1}e^{i\theta})$ for $\theta\in [0,2\pi)$ it is easily seen
that $U(x)$ describes a closed (and simple) contour when $x$ describes the circle $C_{\sqrt{\hat\mu_1/\hat\lambda_1}}$;
moreover, for $x\in C_{\sqrt{\hat\mu_1/\hat\lambda_1}}$, $U(x)$ takes only real values when $x\in\{- \sqrt{\hat\mu_1/\hat\lambda_1},
\sqrt{\hat\mu_1/\hat\lambda_1}\}$.

As a result, we will show that $\chi=0$ if we show that
\begin{equation}
\label{sign}
U\left(-\sqrt{\hat\mu_1/\hat\lambda_1}\right)\times U\left(\sqrt{\hat\mu_1/\hat\lambda_1}\right)>0,
\end{equation}
since (\ref{sign}) will imply that the contour defined by $\{U(x): |x|=\sqrt{\hat\mu_1/\hat\lambda_1}\}$ does not contain the point
$x=0$ in its interior, so that by definition of the index, $\chi=0$.

We have from (\ref{R-A})-(\ref{R-B}) (Hint: $R(x,h(x))=0$ by definition of $h(x))$)
\begin{eqnarray}
A(x,h(x))&=&-\frac{\mu \mu_2}{\alpha} (\lambda(1-h(x))x +\mu_1(x-h(x))
\label{value-A}\\
B(x,h(x))&=&-\frac{\mu \mu_1}{\alpha} (\lambda(1-x)h(x) +\mu_2(h(x)-x)).
\label{value-B}
\end{eqnarray}
Define $x_{-}:=- \sqrt{\hat\mu_1/\hat\lambda_1}$ and $x_{+}:= \sqrt{\hat\mu_1/\hat\lambda_1}$.

By Assumption {\bf A} we know that $x_{-}<-1$ and $x_{+}>1$.
Also note that $h(x_{-})=y_1<1$ and $h(x_{+})=y_2<1$ from Proposition \ref{prop:FayIas}-(d2)
and (\ref{inq-y}). With this, it it is easily seen from (\ref{value-A})-(\ref{value-B}) that
\[
A(x_{-},h(x_{-}))>0 \quad \hbox{and}\quad A(x_{+},h(x_{+}))<0
\]
and
\[
B(x_{-},h(x_{-}))<0 \quad \hbox{and}\quad B(x_{+},h(x_{+}))>0
\]
so that
\[
A(x_{-},h(x_{-}))/B(x_{-},h(x_{-}))<0 \quad \hbox{and}\quad (A(x_{+},h(x_{+}))/B(x_{+},h(x_{+}))<0.
\]
and, therefore,
\begin{equation}
\label{int200}
A(x_{-},h(x_{-}))/B(x_{-},h(x_{-}))\,A(x_{+},h(x_{+}))/B(x_{+},h(x_{+}))>0.
\end{equation}
The above  shows that (\ref{sign}) is true if $r=0$ in the definition of $U(x)$ since in this case
$U(x)=A(x,h(x))/B(x,h(x))$.

Assume that $r=1$ in the definition of $U(x)$ with $x_0<x_+$ and $(\lambda+\mu_1)x_0/(\lambda x_0+\mu_1)\leq
\sqrt{\hat\mu_2/\hat\lambda_2}$.
Since $(x-x_0)<0$ for $x=x_{-}$ and  $(x-x_0)>0$ for $x=x_{+}$ we conclude
from (\ref{int200}) that  $U(x_-)>0$ and $U(x_+)>0$, thereby showing that (\ref{sign}) is also true in this  case.

It remains to investigate the case when $r=1$ with $x_0=x_+$ and $(\lambda+\mu_1)x_0/(\lambda x_0+\mu_1)\leq
\sqrt{\hat\mu_2/\hat\lambda_2}$. Clearly, $U(x_{-})>0$ since, from (\ref{int200}),
$A(x_{-},h(x_{-}))/B(x_{-},h(x_{-}))<0$ and $(x_{-}-x_{0})<0$ because $x_{-}<-1$.

Let us focus on the sign of $U(x_{+})$. We know that the mapping $x\to U(x)$ is continuous for $|x|\leq x_{+}$
and that $U(x_+)\not=0$ when $x_+=x_0$.
Since we have shown that $U(x_{+})>0$ when $x_0<x_+$ and $(\lambda+\mu_1)x_0/(\lambda x_0+\mu_1)\leq
\sqrt{\hat\mu_2/\hat\lambda_2}$, we deduce, by continuity, that necessarily $U(x_+)>0$ when $x_{+}=x_0$ and $(\lambda+\mu_1)x_0/(\lambda x_0+\mu_1)\leq
\sqrt{\hat\mu_2/\hat\lambda_2}$, which concludes the proof.
\hfill\done

\begin{lemma}
\label{lem:Bzero}
Under condition (\ref{stability}) and  Assumption {\bf A},
$B(k(y),y)=0$ for $|y|=1$, $y\not=1$. Also, $B(k(y),y)$ has a zero at $y=1$, with  multiplicity
one.
\end{lemma}

{\bf Proof.}
Fix $|y|=1$, $y\not= 1$. We know from Proposition \ref{prop:FayIas}-(a1) that $|k(y)|<1$.

From (\ref{R-B}) and the fact that $R(k(y),y)=0$ by definition of $k(y)$, we see that
$B(k(y),y)=0$ is equivalent to
\[
0=\lambda (1-k(y))y+\mu_2(y-k(y))=(\lambda(1-k(y))+\mu_2) y - \mu_2 k(y)
\]
that is,
\[
\lambda(1-k(y)+\mu_2) y=\mu_2 k(y).
\]
Taking the absolute value in both sides of the above equation yields
\begin{equation}
\label{inq111}
|\lambda(1-k(y)+\mu_2)|= |\lambda(1-k(y)+\mu_2)y|=|\mu_2 k(y)|<\mu_2.
\end{equation}
But   $|\lambda(1-k(y))+\mu_2)|>\mu_2$ which contradicts (\ref{inq111}). Hence, $B(k(y),y)\not=0$ for
$|y|=1$, $y\not=1$.

Since $k(1)=1$, we see that $B(k(1),1)=B(1,1)=0$ from the definition of $B(x,y)$. Let us show
that the multiplicity of this zero is one. This amounts to showing that
$dB(k(y),y)/dy$ does not vanish at $y=1$.

Differentiating $B(k(y),y)$ w.r.t. $y$ in (\ref{R-B}) (Hint: $R(k(y),y)=0$)
and setting $y=1$, gives
\begin{equation}
\label{inq112}
\frac{d B(k(y),y)}{dy}|_{y=1}=\frac{\mu\mu_1}{\alpha} ((\lambda+\mu_2)k'(1)-\mu_2).
\end{equation}
Let us calculate $k'(1)$, the derivative of $k(y)$ at $y=1$. To this end, let us use (\ref{def-K}) to
differentiate
$R(k(y),y)$ (which is equal to zero) w.r.t. $y$, which gives
\begin{equation}
\label{inq113}
0=\frac{d R(k(y),y)}{dy}|_{y=1}=(\mu\mu_1 -\alpha \lambda_1)k'(1)+\mu \mu_2 -\alpha \lambda_2
\end{equation}
so that $k'(1)=(\alpha \lambda_2-\mu\mu_2)/(\mu\mu_1 -\alpha \lambda_1)$ (note that $\mu\mu_1 -\alpha \lambda_1\not=0$ from
Assumption {\bf A}, which shows that $k'(1)$ is well defined). Plugging this value of $k'(1)$ into (\ref{inq112}) gives
\begin{eqnarray*}
\frac{d B(k(y),y)}{dy}|_{y=1}&=&\frac{\mu\mu_1}{\alpha (\mu\mu_1 -\alpha \lambda_1)}\,
((\alpha \lambda_2-\mu\mu_2)(\lambda+\mu_2)-\mu_2(\mu\mu_1 -\alpha \lambda_1) )\\
&=&\frac{\mu\mu_1}{\alpha(\mu\mu_1 -\alpha \lambda_1)}\alpha (\lambda\lambda_2 +\lambda\mu_2 -\mu\mu_2)\\
&=&\frac{\mu\mu_1}{\mu\mu_1 -\alpha \lambda_1}\mu\mu_2\left(\frac{\lambda\lambda_2}{\mu\mu_2}+\frac{\lambda}{\mu} -1\right)<0
\end{eqnarray*}
under the conditions in (\ref{stability}) (to establish the 2nd equality we have used the definitions of $\alpha$ and $\lambda$). This proves
that $dB(k(y),y)/dy|_{y=1}\not=0$ and completes the proof.
\hfill\done

\newpage
\tableofcontents

\end{document}